\documentclass[preprint,pra,a4paper]{revtex4}%
\usepackage{bm}
\usepackage{amsfonts}
\usepackage{amsmath}
\usepackage{amssymb}
\usepackage{graphicx}%
\providecommand{\U}[1]{\protect\rule{.1in}{.1in}}

\ifx\pdfoutput\relax\let\pdfoutput=\undefined\fi
\newcount\msipdfoutput
\ifx\pdfoutput\undefined\else
\ifcase\pdfoutput\else
\ifx\paperwidth\undefined\else
\ifdim\paperheight=0pt\relax\else\pdfpageheight\paperheight\fi
\ifdim\paperwidth=0pt\relax\else\pdfpagewidth\paperwidth\fi
\fi\fi\fi
\begin{document}
\title{Anomalous States of Positronium}
\author{Chris W Patterson}
\affiliation{Centre for Quantum Dynamics, Griffith University, Nathan QLD 4111 Australia}
\altaffiliation{Guest Scientist, Theoretical Division, Los Alamos National Laboratory, NM
87545 USA}

\begin{abstract}
\textit{The energies and wavefunctions of the two-body Dirac equation for
positronium\ are compared with those of the Pauli approximation and the
Bethe-Salpeter equation. The unusual behavior of the ground-state wavefunction
of the Dirac equation is explained in terms of anomalous bound-state
solutions.}

\end{abstract}
\maketitle

\section{ Introduction}

The two-body Dirac equation for positronium, with only a Coulomb potential,
has been solved numerically by Scott, Schertzer, and Moore \cite{Scott1992}
using very accurate finite element methods. In this paper, their Dirac
energies and wave functions are compared with the results of the Pauli
approximation \cite{Bethe1957} and also with the solutions to the
Bethe-Salpeter equation \cite{Salpeter1951}. The two-body Dirac energies agree
with those of the Pauli approximation to a surprisingly high degree of
accuracy. On the other hand, their ground-state wave functions differ
significantly near the origin. One purpose of this paper is to determine which
ground-state wave function is correct. In order to do so one must consider the
Bethe-Salpeter equation in detail.

In this paper it is shown that both the Dirac equation and the Bethe-Salpeter
equation have two sets of bound-state solutions: the normal `atomic' solutions
and the `anomalous' solutions in which either the electron or the positron is
in a negative-energy state. However, the two-body Dirac equation is not
relativistically invariant and does not properly treat the relative time
between the electron and positron. As a result of this wrong temporal
behavior, the atomic and anomalous states are not orthogonal and can be
coupled by the Coulomb interaction. On the other hand, for the
relativistically invariant Bethe-Salpeter equation, the atomic and anomalous
states are orthogonal because of their different time propagation and there
can be no Coulomb coupling between them. Without this coupling to the
anomalous states, it is shown that there is no unusual behavior near the
origin for the Bethe-Salpeter ground-state wave function. The results in this
paper rely heavily on the previous work of Scott, Schertzer, and Moore from
Ref. \cite{Scott1992}.

In Sec. II a simple derivation for the two-body Dirac equation is given in
both the coordinate and the momentum representation. This simplification uses
recoupling coefficients so that the two-body operators can be readily
evaluated using one-body equations. The results in the coordinate
representation are shown in Appendix A and agree with previous works
\cite{Scott1992},\cite{Malenfant1988}. The results in the momentum
representation are shown in Appendix B and have not been found in the
literature. The momentum representation is useful when calculating the
energies and wave functions of both the Pauli states and the anomalous states.

In Sec. III the Dirac energies for positronium in \cite{Scott1992} are
compared with the Pauli energies and shown to differ by order $mc^{2}%
\alpha^{6}$ or less. However, the Dirac equation also gives anomalous
bound-state solutions for which the radial wave functions are Dirac delta
functions. Unlike the atomic solutions, these anomalous solutions can be
obtained most readily in the momentum representation by using the completeness
relation for the radial wave functions. It is shown that the unusual
ground-state atomic wave function of the Dirac equation is explained by the
Coulomb interaction between the atomic and anomalous states. Moreover, in this
section, the effect of the magnetic potential on the atomic and anomalous
states is derived. This enables one to compare the Dirac energies for the
Coulomb potential alone in Ref. \cite{Scott1992} with the actual positronium
fine structure to order $mc^{2}\alpha^{4}$.

In Sec. IV the Bethe-Salpeter equation for positronium is used to show that
the correct atomic state energies and wave functions are obtained by
eliminating or `projecting out' the anomalous states. Indeed, it is shown that
the correct atomic bound-states occur only when using the Feynman time
propagator $K_{F}$ and the correct anomalous bound-states occur only when
using the retarded time propagator $K_{R}$.

\section{Two-Body Dirac Equations}

The Dirac equation for a free particle is%
\begin{align*}
\boldsymbol{h}_{0}\psi &  =(c\boldsymbol{\alpha}\cdot\boldsymbol{p}%
+mc^{2}\beta)\psi=e\psi,\\
e  &  =\pm\sqrt{p^{2}c^{2}+m^{2}c^{4}}.
\end{align*}
where $\boldsymbol{h}_{0}$ is the Hamiltonian for a particle with energy $e$
and wave function $\psi$. The Dirac matrices $\boldsymbol{\alpha}$ are given
in terms of the Pauli spin matrices $\boldsymbol{\sigma}$ so that
\[
\boldsymbol{\alpha}\boldsymbol{=}\left(
\begin{array}
[c]{cc}%
0 & \boldsymbol{\sigma}\\
\boldsymbol{\sigma} & 0
\end{array}
\right)  ,\ \beta=\left(
\begin{array}
[c]{cc}%
1 & 0\\
0 & -1
\end{array}
\right)  .
\]
For a free electron and positron the two-body Dirac Hamiltonian
$\boldsymbol{H}_{0}$ is the sum of the individual Hamiltonians. The two-body
Dirac equation becomes%
\begin{equation}
\boldsymbol{H}_{0}\Psi=(\boldsymbol{h}_{0}^{e}+\boldsymbol{h}_{0}^{p}%
)\Psi=E\Psi, \label{1.0}%
\end{equation}
or%

\begin{equation}
\boldsymbol{H}_{0}\Psi=(c\boldsymbol{\alpha}_{e}\cdot\boldsymbol{p}_{e}%
+mc^{2}\beta_{e}+c\boldsymbol{\alpha}_{p}\cdot\boldsymbol{p}_{p}+mc^{2}%
\beta_{p})\Psi=E\Psi. \label{1.1}%
\end{equation}
The total energy $E$ is given by the sum of the individual energies for the
electron and positron for which there are four possibilities,
\begin{equation}
E_{\pm\pm}=e_{e}+e_{p}=\pm\sqrt{p_{e}^{2}c^{2}+m^{2}c^{4}}\pm\sqrt{p_{p}%
^{2}c^{2}+m^{2}c^{4}.} \label{1.2}%
\end{equation}
The total wave function $\Psi$ is the direct product of the individual wave
functions,%
\begin{equation}
\Psi=\psi_{e}(\boldsymbol{r}_{e})\times\psi_{p}(\boldsymbol{r}_{p}).
\label{1.3}%
\end{equation}
This direct product wave function has four Dirac components which may be
written in several ways:%
\begin{align*}
\Psi &  =\left(
\begin{array}
[c]{c}%
\psi_{1}(\boldsymbol{r}_{e})\\
\psi_{2}(\boldsymbol{r}_{e})
\end{array}
\right)  \times\left(
\begin{array}
[c]{c}%
\psi_{1}(\boldsymbol{r}_{p})\\
\psi_{2}(\boldsymbol{r}_{p})
\end{array}
\right)  =\binom{_{\psi_{1}(\boldsymbol{r}_{e})\psi_{2}(\boldsymbol{r}_{p}%
)}^{\psi_{1}(\boldsymbol{r}_{e})\psi_{1}(\boldsymbol{r}_{p})}}{_{\psi
_{2}(\boldsymbol{r}_{e})\psi_{2}(\boldsymbol{r}_{p})}^{\psi_{2}(\boldsymbol{r}%
_{e})\psi_{1}(\boldsymbol{r}_{p})}}\equiv\binom{_{\Psi_{12}}^{\Psi_{11}}%
}{_{\Psi_{22}}^{\Psi_{21}}},\\
&  \equiv\Psi_{11}\boldsymbol{e}_{11}+\Psi_{12}\boldsymbol{e}_{12}+\Psi
_{21}\boldsymbol{e}_{21}+\Psi_{22}\boldsymbol{e}_{22},\\
&  \equiv%
\frac12
\{(\Psi_{11}+\Psi_{22})(\boldsymbol{e}_{11}+\boldsymbol{e}_{22})+(\Psi
_{11}-\Psi_{22})(\boldsymbol{e}_{11}-\boldsymbol{e}_{22})\\
&  \ \ \ \ \ \ +(\Psi_{12}+\Psi_{21})(\boldsymbol{e}_{12}+\boldsymbol{e}%
_{21})+(\Psi_{12}-\Psi_{21})(\boldsymbol{e}_{12}-\boldsymbol{e}_{21})\}.
\end{align*}
Here, the latter symmetrized basis is preferred because it has well defined
charge-conjugation and inversion symmetries. With this convention for the
Dirac components $\Psi_{ij}$ of the Dirac vectors $\boldsymbol{e}_{ij},$ the
$i=1,2$ signifies the electron components with positive and negative rest
mass, respectively, and the $j=1,2$ signifies the positron components with
positive and negative rest mass, respectively, such that%
\begin{align*}
mc^{2}\beta_{e}\Psi &  =mc^{2}(\Psi_{11}\boldsymbol{e}_{11}+\Psi
_{12}\boldsymbol{e}_{12}-\Psi_{21}\boldsymbol{e}_{21}-\Psi_{22}\boldsymbol{e}%
_{22}),\\
mc^{2}\beta_{p}\Psi &  =mc^{2}(\Psi_{11}\boldsymbol{e}_{11}-\Psi
_{12}\boldsymbol{e}_{12}+\Psi_{21}\boldsymbol{e}_{21}-\Psi_{22}\boldsymbol{e}%
_{22}).
\end{align*}

When the electron and positron interact with a Coulomb potential, it is useful
to transform to the relative coordinates,
\begin{equation}
\boldsymbol{\rho}=\boldsymbol{r}_{e}-\boldsymbol{r}_{p},\ \ \ \boldsymbol{R}=%
\frac12
(\boldsymbol{r}_{e}+\boldsymbol{r}_{p}), \label{1.4}%
\end{equation}
and their conjugate momenta%
\begin{equation}
\boldsymbol{\pi}\equiv-i\hbar\boldsymbol{\nabla}_{\rho}=%
\frac12
(\boldsymbol{p}_{e}-\boldsymbol{p}_{p}),\ \ \ \boldsymbol{P}\equiv
-i\hbar\boldsymbol{\nabla}_{R}=\boldsymbol{p}_{e}+\boldsymbol{p}_{p}.
\label{1.5}%
\end{equation}
Letting the total momentum be zero, $\boldsymbol{P=0,}$ corresponding to the
center of momentum frame, one finds%
\[
\boldsymbol{p}_{e}+\boldsymbol{p}_{p}=\boldsymbol{0},\ \ \ \boldsymbol{p}%
_{e}=-\boldsymbol{p}_{p}=\boldsymbol{\pi},
\]
and the two-body Dirac equation (\ref{1.1}) for free particles becomes%
\begin{equation}
\boldsymbol{H}_{0}\Psi=(c\boldsymbol{\alpha}_{e}\cdot\boldsymbol{\pi}%
+mc^{2}\beta_{e}-c\boldsymbol{\alpha}_{p}\cdot\boldsymbol{\pi}+mc^{2}\beta
_{p})\Psi=E\Psi. \label{1.6}%
\end{equation}
\qquad

From (\ref{1.2}) the free particle energies are then,%
\begin{equation}
E_{++}=+2e_{0},\ E_{--}=-2e_{0},\ \ \ E_{+-}=E_{-+}=0, \label{1.7a}%
\end{equation}
where%
\begin{equation}
e_{0}=\sqrt{\pi^{2}c^{2}+m^{2}c^{4}.} \label{1.7b}%
\end{equation}
\newline The Dirac free particle states $\Psi_{++}$ and $\Psi_{--}$,
corresponding to energies $E_{++}$ and $E_{--}$, respectively, comprise the
particle and antiparticle states of `$atomic$' positronium for the Dirac
`hole' theory. The free particle states $\Psi_{+-}$ and $\Psi_{-+}$,
corresponding to energy $E=0$, are degenerate for all relative momentum $\pi$
and give rise to the `$anomalous$' states. Note that this degeneracy only
occurs for equal masses where $m_{e}=m_{p}=m$ and not for hydrogenic atoms in
general. These anomalous degenerate states will be strongly coupled by the
electric and magnetic potential which will split their degeneracy. Thus the
weak electromagnetic potential has a very strong effect on these degenerate
states. One of the main results of this paper is to explicitly find the
energies and wave functions of the anomalous states resulting from the
splitting of this degeneracy and to calculate the interaction of these
anomalous states with the atomic ground-state. Indeed, as explained in the
next section, it is this interaction which explains the unusual behavior of
the Dirac ground-state near the origin.

For a Coulomb potential, $V_{C}(\rho)$, it is useful to use spherical
coordinates, $\boldsymbol{\rho}\equiv(\rho,\theta,\varphi)$, where
\begin{equation}
V_{C}(\rho)=-e^{2}/\rho, \label{1.16}%
\end{equation}
and the two-body Dirac equation (\ref{1.6}) becomes%
\begin{equation}
(c\boldsymbol{\alpha}_{e}\cdot\boldsymbol{\pi}+mc^{2}\beta_{e}%
-c\boldsymbol{\alpha}_{p}\cdot\boldsymbol{\pi}+mc^{2}\beta_{p})\Psi
=(E+e^{2}/\rho)\Psi. \label{1.17}%
\end{equation}

One can couple the individual spin functions $\chi_{\sigma_{e}}^{%
\frac12
}$ and $\chi_{\sigma_{p}}^{%
\frac12
}$ of the electron and positron to find functions of total spin $S$ and
projection $S_{z}=\Sigma$ such that%
\begin{equation}
\Omega_{\Sigma}^{S}\equiv\lbrack\chi_{e}^{%
\frac12
}\chi_{p}^{%
\frac12
}]_{\Sigma}^{S}=\sum\limits_{\sigma_{e},\sigma_{p}}C_{\sigma_{e}%
\text{\ }\sigma_{p}\text{ }\Sigma}^{%
\frac12
\text{ \ }%
\frac12
\text{ \ }S}\chi_{\sigma_{e}}^{%
\frac12
}\chi_{\sigma_{p}}^{%
\frac12
}, \label{1.8}%
\end{equation}
where $C_{\sigma_{e}\text{\ }\sigma_{p}\text{ }\Sigma}^{%
\frac12
\text{ \ }%
\frac12
\text{ \ }S}$ are the coupling coefficients for spin-1/2 particles.
Explicitly, the four possible spin functions $\Omega_{\Sigma}^{S}$ for $S=0$
and $S=1$ are%

\begin{align*}
\Omega_{0}^{0}  &  \equiv\lbrack\chi_{e}^{%
\frac12
}\chi_{p}^{%
\frac12
}]_{0}^{0}=[\chi_{%
\frac12
}^{%
\frac12
}\chi_{-%
\frac12
}^{%
\frac12
}-\chi_{-%
\frac12
}^{%
\frac12
}\chi_{%
\frac12
}^{%
\frac12
}]/\sqrt{2},\\
\Omega_{0}^{1}  &  \equiv\lbrack\chi_{e}^{%
\frac12
}\chi_{p}^{%
\frac12
}]_{0}^{1}=[\chi_{%
\frac12
}^{%
\frac12
}\chi_{-%
\frac12
}^{%
\frac12
}+\chi_{-%
\frac12
}^{%
\frac12
}\chi_{%
\frac12
}^{%
\frac12
}]/\sqrt{2},\\
\Omega_{-1}^{1}  &  \equiv\lbrack\chi_{e}^{%
\frac12
}\chi_{p}^{%
\frac12
}]_{-1}^{1}=\chi_{-%
\frac12
}^{%
\frac12
}\chi_{-%
\frac12
}^{%
\frac12
},\\
\Omega_{1}^{1}  &  \equiv\lbrack\chi_{e}^{%
\frac12
}\chi_{p}^{%
\frac12
}]_{1}^{1}=\chi_{%
\frac12
}^{%
\frac12
}\chi_{%
\frac12
}^{%
\frac12
}.
\end{align*}
Using the exchange symmetry of the spin functions, one finds%
\begin{equation}
\lbrack\chi_{e}^{%
\frac12
}\chi_{p}^{%
\frac12
}]_{\Sigma}^{S}=(-1)^{S+1}[\chi_{p}^{%
\frac12
}\chi_{e}^{%
\frac12
}]_{\Sigma}^{S}. \label{1.9}%
\end{equation}
So, under particle exchange, the $S=0$ spin functions are antisymmetric and
the $S=1$ spin functions are symmetric.

Because the total angular momentum $J$ is conserved for a Coulomb potential,
the spherical harmonics for orbital angular momentum $Y_{M}^{L}(\theta
,\varphi)$ are coupled with the total spin functions to obtain eigenfunctions
of total angular momentum $J$ and projection $J_{z}=N$ where%
\begin{equation}
\lbrack Y^{L}(\theta,\varphi)\Omega^{S}]_{N}^{J}=\sum_{M,\Sigma}%
C_{M\text{\ }\Sigma\text{ }N}^{L\text{ \ }S\text{ \ }J}Y_{M}^{L}%
(\theta,\varphi)\Omega_{\Sigma}^{S}. \label{1.10}%
\end{equation}
\ There are four possible angular momentum states $[Y^{L}\Omega^{S}]_{N}^{J}$
for a given $J$ and $N$ depending on the spin $S$ and the orbital angular
momentum $L$, namely,%
\begin{equation}
\ [Y^{J+1}\Omega^{1}]_{N}^{J}=[Y^{J+1}[\chi_{e}^{%
\frac12
}\chi_{p}^{%
\frac12
}]^{1}]_{N}^{J}=[[Y^{J+1}\chi_{e}^{%
\frac12
}]^{J+%
\frac12
}\chi_{p}^{%
\frac12
}]_{N}^{J}, \label{1.11}%
\end{equation}%
\[
\ [Y^{J-1}\Omega^{1}]_{N}^{J}=[Y^{J-1}[\chi_{e}^{%
\frac12
}\chi_{p}^{%
\frac12
}]^{1}]_{N}^{J}=[[Y^{J-1}\chi_{e}^{%
\frac12
}]^{J-%
\frac12
}\chi_{p}^{%
\frac12
}]_{N}^{J}.
\]%
\[
\ [Y^{J}\Omega^{0}]_{N}^{J}=[Y^{J}[\chi_{e}^{%
\frac12
}\chi_{p}^{%
\frac12
}]^{0}]_{N}^{J}=a[[Y^{J}\chi_{e}^{%
\frac12
}]^{J+%
\frac12
}\chi_{p}^{%
\frac12
}]_{N}^{J}-b[[Y^{J}\chi_{e}^{%
\frac12
}]^{J-%
\frac12
}\chi_{p}^{%
\frac12
}]_{N}^{J},\text{\ }%
\]%
\[
~[Y^{J}\Omega^{1}]_{N}^{J}=[Y^{J}[\chi_{e}^{%
\frac12
}\chi_{p}^{%
\frac12
}]^{1}]_{N}^{J}=b[[Y^{J}\chi_{e}^{%
\frac12
}]^{J+%
\frac12
}\chi_{p}^{%
\frac12
}]_{N}^{J}+a[[Y^{J}\chi_{e}^{%
\frac12
}]^{J-%
\frac12
}\chi_{p}^{%
\frac12
}]_{N}^{J},
\]
where%
\begin{equation}
a=\sqrt{\frac{J+1}{2J+1}},\ \ \ b=\sqrt{\frac{J}{2J+1}} \label{1.12}%
\end{equation}
are the angular momentum recoupling coefficients for spin-1/2 particles.
\ Making use of the spin exchange symmetry (\ref{1.9}), it follows that we can
also write (\ref{1.11}) as%

\begin{align}
\lbrack Y^{J+1}\Omega^{1}]_{N}^{J}  &  =[[Y^{J+1}\chi_{p}^{%
\frac12
}]^{J+%
\frac12
}\chi_{e}^{%
\frac12
}]_{N}^{J},\label{1.13}\\
\lbrack Y^{J-1}\Omega^{1}]_{N}^{J}  &  =[[Y^{J-1}\chi_{p}^{%
\frac12
}]^{J-%
\frac12
}\chi_{e}^{%
\frac12
}]_{N}^{J},\nonumber\\
\lbrack Y^{J}\Omega^{0}]_{N}^{J}  &  =-a[[Y^{J}\chi_{p}^{%
\frac12
}]^{J+%
\frac12
}\chi_{e}^{%
\frac12
}]_{N}^{J}+b[[Y^{J}\chi_{p}^{%
\frac12
}]^{J-%
\frac12
}\chi_{e}^{%
\frac12
}]_{N}^{J},\nonumber\\
\lbrack Y^{J}\Omega^{1}]_{N}^{J}  &  =b[[Y^{J}\chi_{p}^{%
\frac12
}]^{J+%
\frac12
}\chi_{e}^{%
\frac12
}]_{N}^{J}+a[[Y^{J}\chi_{p}^{%
\frac12
}]^{J-%
\frac12
}\chi_{e}^{%
\frac12
}]_{N}^{J}.\nonumber
\end{align}

\subsection{Coordinate Representation\qquad}

In the coordinate representation the single particle operator
$\boldsymbol{\sigma\cdot\pi}$ acting on the radial functions $y(\rho)$ is
given by \cite{Bethe1957},\cite{Sakurai1967}%

\begin{align}
\boldsymbol{\sigma\cdot\pi}\frac{y}{\rho}[Y^{J}\chi^{%
\frac12
}]_{N}^{J-%
\frac12
}  &  =\frac{i\hbar}{\rho}(\frac{dy}{d\rho}+\frac{Jy}{\rho})[Y^{J-1}\chi^{%
\frac12
}]_{N}^{J-%
\frac12
},\label{1.14}\\
\boldsymbol{\sigma\cdot\pi}\frac{y}{\rho}[Y^{J}\chi^{%
\frac12
}]_{N}^{J+%
\frac12
}  &  =\frac{i\hbar}{\rho}(\frac{dy}{d\rho}-\frac{(J+1)y}{\rho})[Y^{J+1}\chi^{%
\frac12
}]_{N}^{J+%
\frac12
}.\nonumber
\end{align}
Using the recoupling (\ref{1.11})-(\ref{1.13}) and the single particle
operator equations (\ref{1.14}), it is now a simple matter to evaluate
$\boldsymbol{\alpha}_{e}\cdot\boldsymbol{\pi}$ and $-\boldsymbol{\alpha}%
_{p}\cdot\boldsymbol{\pi}$ in (\ref{1.17}) and to derive the two-body Dirac
partial differential equations in the coordinate representation. Note that
$y(\rho)$ satisfies the boundary condition $y(\rho)=0$. The resulting three
sets of equations for free particles are given in Appendix A. These equations
agree with previous works \cite{Malenfant1988}, \cite{Scott1992} where cases
1A, 2A, and 3A correspond to the sets 1, 3, and 2 in Ref. \cite{Scott1992}.

\subsection{Momentum Representation}

Calculations can also be performed in the momentum representation formed by
`Fourier analysis' of the equations in the coordinate representation using
spherical Bessel functions $N_{Jm}j_{J}(k_{m}\rho)$ where $\pi=\hbar k_{m}$.
The $k_{m}$ are determined by the boundary condition%
\begin{equation}
j_{J}(k_{m}\rho_{0})=0,\text{ for }m=1,2,..., \label{1.15}%
\end{equation}
with $\rho_{0}\gg2na_{0\text{ }}$ (where $a_{0}$ is the Bohr radius). The
normalizations $N_{Jm}$ are determined by
\[
N_{Jm}\int_{0}^{\rho_{0}}\rho^{2}j_{J}^{2}(k_{m}\rho)d\rho=1.
\]
\ The Bessel functions $j_{L}(k_{m}\rho)$ for different $L=J-1,J,J+1$ have the
same normalization $N_{Jm}$ independent of $L$ given by%
\[
N_{Jm}=\sqrt{\frac{2}{\rho_{0}^{3}j_{J\pm1}^{2}(k_{m}\rho_{0})}},
\]
where $j_{J+1}(k_{m}\rho_{0})=-j_{J-1}(k_{m}\rho_{0}).$ For high $m\gg1$ where
$k_{m}\rho_{0}\gg1$ one has the approximations%

\begin{align*}
N_{Jm}  &  \simeq k_{m}\sqrt{2/\rho_{0}},\\
k_{m}  &  \simeq(%
\frac12
J\pi+m\pi)/\rho_{0}.
\end{align*}
The functions $\rho j_{L}(k_{m}\rho)$ form an orthonormal set such that%
\[
N_{Jm}^{2}\int_{0}^{\rho_{0}}\rho^{2}j_{L}(k_{m}\rho)j_{L}(k_{n}\rho
)d\rho=\delta_{mn}\text{ for }L=J-1,\text{ }J,\text{ }J+1,
\]
where the $k_{m}$ are, again, determined solely by (\ref{1.15}) (for $L=J$%
)$.$\qquad

As in the coordinate basis (\ref{1.11}), there are four different momentum
bases $|L,S,k\rangle$ for a given $J,N,$ and $k$:%

\begin{align}
\ |J+1,1,k\rangle &  =iN_{Jk}\rho j_{J+1}(k\rho)[Y^{J+1}\Omega^{1}]_{N}%
^{J},\label{1.18}\\
|J-1,1,k\rangle &  =iN_{Jk}\rho j_{J-1}(k\rho)[Y^{J-1}\Omega^{1}]_{N}%
^{J}.\nonumber\\
\ \ \ \ \ |J,0,k\rangle &  =N_{Jk}\rho j_{J}(k\rho)[Y^{J}\Omega^{0}]_{N}%
^{J},\nonumber\\
\ \ \ \ \ \ \ |J,1,k\rangle &  =N_{Jk}\rho j_{J}(k\rho)[Y^{J}\Omega^{1}%
]_{N}^{J}.\nonumber
\end{align}
Note that the scale factor $\rho$ for spherical coordinates is used in the
definition of the wave function as it was in the coordinate representation
(\ref{1.14}) so that these wave functions are also zero at $\rho=0$. When
recoupling, it is useful to define linear combinations of $|J+1,1,k\rangle$
and $|J-1,1,k\rangle$, namely,%
\begin{align}
|J\alpha,1,k\rangle &  \equiv a|J+1,1,k\rangle+b|J-1,1,k\rangle,\label{1.19}\\
|J\beta,1,k\rangle &  \equiv-b|J+1,1,k\rangle+a|J-1,1,k\rangle,\nonumber
\end{align}
where $a$ and $b$ are the recoupling coefficients given in (\ref{1.12}).
\ Using the one-particle equations for the spherical Bessel functions, one can
find the equivalent equations to (\ref{1.14}) for the momentum representation:%
\begin{align}
\boldsymbol{\sigma\cdot\pi\{}j_{J}(k\rho)[[Y^{J}\chi^{%
\frac12
}]_{N}^{J-%
\frac12
}\}  &  =i\hbar k\{j_{J-1}(k\rho)[[Y^{J-1}\chi^{%
\frac12
}]_{N}^{J-%
\frac12
}\},\label{1.20}\\
\boldsymbol{\sigma\cdot\pi\{}j_{J}(k\rho)[[Y^{J}\chi^{%
\frac12
}]_{N}^{J+%
\frac12
}\}  &  =-i\hbar k\{j_{J+1}(k\rho)[[Y^{J+1}\chi^{%
\frac12
}]_{N}^{J+%
\frac12
}\}.\nonumber
\end{align}
Recoupling the angular wave functions as in (\ref{1.11})-(\ref{1.13}), one can
evaluate $\boldsymbol{\sigma}_{e}\boldsymbol{\cdot}$\textbf{$\pi$} and
$-\boldsymbol{\sigma}_{p}\boldsymbol{\cdot}$\textbf{$\pi$} in (\ref{1.17})
when operating on the wave functions in (\ref{1.18}). The three sets of
equations for free particles in momentum space in cases 1B, 2B, and 3B,
analogous to cases 1A, 2A, and 3A, can now be derived and solved analytically
as shown in Appendix B. The free particle wave functions in (\ref{1.18}),
(\ref{1.19}) may also be expanded in terms of products of their single
particle components $g_{n_{e}}^{\ell_{e}\text{ }j_{e}}(kr_{e},\theta
_{e},\varphi_{e})g_{n_{p}}^{\ell_{p}\text{ }j_{p}}(kr_{p},\theta_{p}%
,\varphi_{p}),$ where%
\[
g_{n}^{\ell\text{ }j}(kr,\theta,\varphi)\equiv j_{\ell}(kr)[Y^{\ell}%
(\theta,\varphi)\chi^{%
\frac12
}]_{n}^{j},
\]
as shown in Appendix B.

\section{Dirac and Pauli Solutions for Positronium}

For a given case of angular momentum states $[Y^{L}\Omega^{S}]^{J}$ shown in
Appendix B, the Dirac solutions with a Coulomb potential arise from
combinations of the four free particle basis $\Psi_{++},\ \Psi_{--},\ \Psi
_{S},$ $\Psi_{A}$ which have the same $C$ parity from charge-conjugation
symmetry and $P$ parity from inversion symmetry as shown by Malenfant
\cite{Malenfant1988}. Two are the atomic solutions which are corresponding to
the $\Psi_{++}$ and $\Psi_{--}$ states. Indeed, the $\Psi_{++}$ states
correspond to the Pauli atomic states of positronium and the $\Psi_{--}$
states correspond to the Pauli atomic antiparticle states with
negative-energy. The remaining two symmetrized states $\Psi_{S}=(\Psi_{+-}+$
$\Psi_{-+})/\sqrt{2}$ and $\Psi_{A}=(\Psi_{+-}-$ $\Psi_{-+})/\sqrt{2}$
correspond to the anomalous states.

\subsection{Coulomb Potential}

Appendix A and B are equations for free two-particle Dirac wave functions in
the relative coordinates. For a Coulomb potential (\ref{1.16}), changes must
be made to the equations in Appendices A and B. In Appendix A one must change
$E$ to $E-V_{C}=E+e^{2}/\rho$. In Appendix B one must change $E$ to
$E\delta_{kk^{\prime}}-V_{kk^{\prime}}^{i}$ using the appropriate combination
of orthonormal spherical Bessel functions $j_{L}(k\rho)$ to evaluate the
matrix elements $V_{kk^{\prime}}^{i}$ as described in this Appendix, depending
on the case 1B, 2B, or 3B.

\subsubsection{Atomic States}

In this section, the finite element calculations for the Dirac energies
$E_{D}$ of \cite{Scott1992} in the coordinate representation are compared to
the analytic results of the Pauli approximation $E_{P}$, valid to order
$mc^{2}\alpha^{4}$ for the Coulomb energies of the positronium atom$.$ The
Pauli basis only includes the $\Psi_{++}$ free particle states as shown in
Appendix B. The development of Bethe and Salpeter \cite{Bethe1957} is used for
the Coulomb energies $E_{P}$ of the Pauli approximation adopting their
notation. \ While Bethe and Salpeter give the results for both the Coulomb and
the Breit interaction together (as derived previously by Ferrell
\cite{Ferrell1951}), one can readily extract only the Coulomb part from their
results. Accordingly, one has the following formulas for the Pauli
approximation to the Coulomb energies $E_{P}$ for a given $(nLSJ$ $)$ state of
positronium:%
\begin{equation}
E_{P}=2mc^{2}+H_{0}+H_{1}+H_{3}^{C}+H_{4}^{C}. \label{1.26}%
\end{equation}
where
\begin{align}
H_{0}  &  =-\frac{mc^{2}\alpha^{2}}{4n^{2}},\label{1.25}\\
H_{1}  &  =\frac{3mc^{2}\alpha^{4}}{64n^{4}}-\frac{mc^{2}\alpha^{4}}%
{8n^{3}(2L+1)},\nonumber\\
H_{3}^{C}  &  =\frac{mc^{2}\alpha^{4}}{8n^{3}L(L+1)(2L+1)}\left\{
\begin{tabular}
[c]{c}%
$L$\\
$-1$\\
$-(L+1)$%
\end{tabular}%
\begin{tabular}
[c]{c}%
$\text{ }J=L+1$\\
$\text{ }J=L$\\
$\text{ }J=L-1$%
\end{tabular}
\ \ \ \ \ \ \ \ \ \ \ \ \right\}  \text{ }(1-\delta_{L0})\delta_{S1}%
,\nonumber\\
H_{4}^{C}  &  =\frac{mc^{2}\alpha^{4}}{8n^{3}}\delta_{L0}.\nonumber
\end{align}

The term $H_{0}$ is just the second order positronium energy from the
Schrodinger equation and $H_{1}$ is the relativistic kinetic correction to the
fourth order. The terms $H_{3}^{C}+H_{4}^{C}$ correspond to the fine structure
corrections of positronium for a Coulomb potential to the fourth order . Note
the Kronecker deltas are such that $H_{3}^{C}=0$ if $L=0$ or $S=0$. Also, note
that $H_{4}^{C}$ cancels the second term of $H_{1}$ when $L=0$. Later, the
magnetic terms $H_{2}^{B},$ $H_{3}^{B},$ and $H_{5}^{B}$ are considered for
the Breit potential which also contribute to the fourth order.

The Dirac energies, $E_{D}-2mc^{2},$ in Ref. \cite{Scott1992} for all $(nLSJ)$
states up to $n=3$ are compared with the Pauli energies to fourth order,
$E_{P}-2mc^{2},$ in Table I where cases 1,2,3 correspond to sets 1,3,2 in Ref.
\cite{Scott1992}. All calculations use the approximate value $\alpha=1/137$ as
in Ref. \cite{Scott1992} with the energies in units of $Hartree=mc^{2}%
\alpha^{2}$ and the differences in energies $E_{D}-E_{P}$ is in units of $Mhz$
using the conversion $Hartree=(mc^{2}\alpha^{2}/h)Hz=6.579684\times10^{9}$
$MHz$. The agreement in Table 1 between $E_{D}$ and $E_{P}$ is surprising. It
has been shown by Ishidzu \cite{Ishidzu1951a,Ishidzu1951b} that the energies
for the atomic states for the two-body Dirac equation in a Coulomb potential
can be expanded in a power series in $\alpha^{2}.$ Thus, according to
Ishidzu,\ the energy differences $E_{D}-E_{P}$ in Table 1 should be of order
$mc^{2}\alpha^{6}/h\sim20\ MHz$ which is approximately the agreement found for
the $L=0$ states.

Actually, one can fit the energy differences $E_{D}-E_{P}$ for the three $L=0$
states $(n000)$ to the expression $\nu_{0}/n^{3}$ with $\nu_{0}=-10.6376$
$Mhz$ and a standard deviation of only $\sigma=0.5\ Khz.$ Similarly, one can
fit the energy differences $E_{D}-E_{P}$ for the three $L=0$ states $(n011)$
to the expression $\nu_{1}/n^{3}$ with $\nu_{1}=-7.2724$ $Mhz$ and a standard
deviation of $\sigma=8\ Khz.$ The $L=1$ states have accuracies of \ $\sim10$
$kHz$ or less and the $L=2$ states have accuracies of $\sim1$ $kHz$ or less.
The overall standard deviation for all 18 $(nLSJ)$ states is then only
$\sigma=7\ Khz$ using the constants $\nu_{0}$ and $\nu_{1}$ . Altogether, this
remarkable agreement is proof of the accuracy of the calculations in Ref.
\cite{Scott1992} and corresponds to 16 significant figures with respect to the
positronium rest mass $2mc^{2}$. Note that this numerical accuracy is much
greater than either the theoretical accuracy of QED calculations for atomic
positronium, presently to order $mc^{2}\alpha^{6},$ or the experimental error
in any positronium spectroscopy, both of which are typically of $\sim1\ MHz$ accuracy.

However, this numerical accuracy in the energy differences $E_{D}-E_{P}$ is
somewhat misleading in that the Dirac equation with a Coulomb potential is
missing all terms $mc^{2}\alpha^{n}$ of odd order $n$%
\ \cite{Ishidzu1951a,Ishidzu1951b}, the largest of which is $mc^{2}\alpha
^{5},$ and $E_{P}$ is only calculated to order $mc^{2}\alpha^{4}$. In fact, it
has been shown by Fulton and Martin \cite{Fulton1954} that there is an energy
shift for the $L=0$ states of positronium, due to a Coulomb term of order
$mc^{2}\alpha^{5}$ in the Pauli approximation, given approximately by%
\[
\Delta E_{L=0}\simeq\frac{-mc^{2}\alpha^{5}}{8n^{3}}=-\frac{320}{n^{3}}Mhz.
\]
Thus, for a Coulomb potential, the Pauli energy of the ground-state is
approximately $320\ Mhz$ lower than the Dirac energy. There are many more
terms of this order due to QED corrections for the $(nLSJ)$ states as treated
in Ref. \cite{Fulton1954}. Finally, there are the magnetic terms of order
$mc^{2}\alpha^{4}$, considered below, which contribute to the fine structure.%

\begin{table}[tbp] \centering
\begin{tabular}
[c]{|l|c|l|l|l|}\hline
$n$ $L$ $S$ $J$ & $Case$ & $E_{D}-2mc^{2}$ $(Hartree)$ & $E_{P}-2mc^{2}$
$(Hartree)$ & $E_{D}-E_{P}$ $(MHz)$\\\hline
1 0 0 0 & 1 & -0.249 997 504 147 52 & \multicolumn{1}{|l|}{-0.249 997 502 530
77} & -10.6377\\
1 0 1 1 & 3 & -0.249 997 503 636 33 & \multicolumn{1}{|l|}{-0.249 997 502 530
77} & \ -7.2742\\
2 0 0 0 & 1 & -0.062 499 844 110 14 & \multicolumn{1}{|l|}{-0.062 499 843 908
17} & \ -1.3289\\
2 0 1 1 & 3 & -0.062 499 844 044 37 & -0.062 499 843 908 17 & \ -0.8962\\
2 1 0 1 & 1 & -0.062 500 121 403 76 & \multicolumn{1}{|l|}{-0.062 500 121 404
75} & \ \ 0.0066\\
2 1 1 0 & 3 & -0.062 500 398 904 81 & -0.062 500 398 901 33 & \ -0.0229\\
2 1 1 1 & 2 & -0.062 500 260 153 03 & \multicolumn{1}{|l|}{-0.062 500 260 153
04} & \ \ 0.0001\\
2 1 1 2 & 3 & -0.062 499 982 656 70 & -0.062 499 982 656 46 & \ -0.0016\\
3 0 0 0 & 1 & -0.027 777 747 004 71 & \multicolumn{1}{|l|}{-0.027 777 746 944
82} & \ -0.3941\\
3 0 1 1 & 3 & -0.027 777 746 985 12 & -0.027 777 746 944 82 & \ -0.2651\\
3 1 0 1 & 1 & -0.027 777 829 165 71 & \multicolumn{1}{|l|}{-0.027 777 829 166
03} & \ \ 0.0021\\
3 1 1 0 & 3 & -0.027 777 911 388 16 & -0.027 777 911 387 24 & \ -0.0060\\
3 1 1 1 & 2 & -0.027 777 870 276 58 & \multicolumn{1}{|l|}{-0.027 777 870 276
64} & \ \ 0.0004\\
3 1 1 2 & 3 & -0.027 777 788 055 52 & -0.027 777 788 055 43 & \ -0.0006\\
3 2 0 2 & 1 & -0.027 777 796 277 54 & \multicolumn{1}{|l|}{-0.027 777 796 277
55} & \ \ 0.0001\\
3 2 1 1 & 3 & -0.027 777 820 944 03 & -0.027 777 820 943 91 & \ -0.0007\\
3 2 1 2 & 2 & -0.027 777 804 499 68 & -0.027 777 804 499 67 & \ -0.0001\\
3 2 1 3 & 3 & -0.027 777 779 833 31 & \multicolumn{1}{|l|}{-0.027 777 779 833
31} & \ -0.0000\\\hline
\end{tabular}
\caption{Comparison of Two-Body Dirac Energies of Scott, et al. with the Pauli Energies for a Coulomb
Potential}\label{Table 1}%
\end{table}%
\qquad$\frac{\frac{{}}{{}}}{{}}$

The Dirac radial components of the ground-state wave function, shown in Fig.
1, are calculated in the coordinate representation using equations
(\ref{A.1b}) for Case 1A. As shown in Ref. \cite{Scott1992}, accurate
ground-state wave functions for the Dirac equation can only be achieved with a
grid which has elements in the region of $\rho\sim1$ $Fermi$. \ In this work,
a grid is used to cover the three Regions shown in Fig. 1: Region 1, with 10
equal elements of $\Delta\rho=10^{-5}$ $Bohr$, Region 2, with 9 equal elements
of $\Delta\rho=0.002$ $Bohr$ and Region 3, with 79 equal elements of
$\Delta\rho=0.5$ $Bohr$ for $n=1$ states. The endpoint and element size of
this last region varies proportionately with $n$ of the positronium radius
$a=2n$~$Bohr$. With five point Lagrangian interpolations, the 100 finite
elements correspond to 399 grid points. In atomic units, Region 1 is near the
classical radius of the electron or positron $\rho=\alpha^{2}\sim$
$(1/137)^{2}$ $Bohr$. Region 2 is near the Compton wavelength $\rho=\alpha$
$\sim(1/137)$ $Bohr$ and Region 3 is near the radius $\rho=2n$ $Bohr$ of
positronium. In this figure, a log-log scale is used to investigate the wave
function at very small $\rho$. The large dots in the figure correspond to the
grid points used for the three different regions in the Dirac calculations.
For five point Lagrangian interpolation, there are four grid points for every
element. These dots merge into a continuous line when the spacing becomes
small on this log scale.

The Pauli wave functions, calculated using the $\Psi_{++}$ basis in Appendix
B, are also shown in Fig. 1. One might expect very good agreement between the
Dirac and Pauli wave functions as is the case for their energies. However, the
Dirac solution for $y_{11}^{2}$ and $y_{22}^{2}$ differ greatly from their
Pauli solutions in Region 1$.$ This unusual behavior is the result of the
coupling of atomic and anomalous states by the Coulomb potential. As shown in
Fig. 1, the Dirac wave function $y_{11}$ for the positronium ground-state is
approximately the Schrodinger ground-state wave function $y_{S}$ such that
$y_{11}^{2}\simeq y_{S}^{2}=%
\frac12
\rho^{2}e^{-\rho}$ in atomic units$.$ Also shown in Fig. 1, the Pauli wave
function $y_{22}^{2}$ reaches its maximum near Region 2. On the other hand,
the Dirac wave function $y_{22}^{2}$ reaches its maximum near Region 1 as has
been shown previously in Ref. \cite{Scott1992}. In order to explain the
unusual behavior of the Dirac ground-state solutions in Region 1, it is
necessary to first consider the anomalous states arising from the $\Psi_{+-}$
and $\Psi_{-+}$ free particle wave functions.

\subsubsection{Anomalous States and the Discrete Variable Representation
(DVR)}

It is shown below, both numerically and analytically, that there are
bound-state solutions of the Dirac equation for the anomalous states. The
anomalous states can be obtained most readily in the momentum basis using
Appendix B because this basis allows one to clearly distinguish the anomalous
states $\Psi_{+-}$ and $\Psi_{-+}$ from the atomic states $\Psi_{++}$ and
$\Psi_{--}$. However, first we treat the anomalous states in the coordinate basis.

In the coordinate basis, the finite element calculations using (\ref{A.1b})
for the angular momentum states $[Y^{0}\Omega^{0}]^{0}$ give solutions for
both the Dirac atomic bound-states and the Dirac anomalous bound-states
simultaneously. The energy solutions for the anomalous states in Region 1 are
shown in Fig.\ 2 and their respective wave function solutions are shown in
Fig.\ 3. The solutions shown are bound-states.

As seen in Figs. 2 and 3, the solutions are approximately Dirac delta
functions,
\begin{equation}
(y_{11}^{0}-y_{22}^{0})/\sqrt{2}\simeq C\delta(\rho-\rho_{i}), \label{2.8}%
\end{equation}
with energy $E_{i}\simeq-e^{2}/\rho_{i}$ where $C$ is the normalization and
\begin{equation}
\rho_{i}=i\Delta\rho\label{2.9}%
\end{equation}
are at the Lagrangian nodes themselves with spacing $\Delta\rho=10^{-4}/40$.
For clarity only every fourth wave function is shown in Fig.\ 3, corresponding
to the vertical lines in Fig.~2 at the element boundaries. These delta
functions are approximately the Lagrangian nodal functions themselves although
they have been orthogonalized by the diagonalization and are now centered at
the nodes. Note that the delta functions in the coordinate bases for the
anomalous states are strictly linked to the grid points. This is not an
artefact of the numerical calculations in the coordinate bases as demonstrated
below using the momentum bases. In Fig. 3 the component $(y_{12}^{+}%
+y_{21}^{+})$ is too small to be seen.

Instead of solving the set of equations (\ref{A.1b}), these same anomalous
bound-state solutions can be obtained more directly by solving the simpler
equation in the same finite element coordinate basis for Case~1A corresponding
to
\[
(y_{11}^{0}-y_{22}^{0})=\frac{2mc^{2}(y_{11}^{0}+y_{22}^{0})}{E+e^{2}/\rho},
\]
where $E$ has been replaced by $E+e^{2}/\rho$. The radial function
$(y_{11}^{0}-y_{22}^{0})$ is singular near $E=-e^{2}/\rho$ unless $(y_{11}%
^{0}+y_{22}^{0})=0$ in which case one obtains the simple equation%
\begin{equation}
-(e^{2}/\rho)(y_{11}^{0}-y_{22}^{0})=E(y_{11}^{0}-y_{22}^{0}). \label{x.1}%
\end{equation}
The anomalous energies and wave functions are found by simply diagonalizing
the potential matrix of $V_{C}(\rho)=-e^{2}/\rho$ in the coordinate basis. The
resulting radial solutions of (\ref{x.1}) are indistinguishable from the
solutions shown in Fig. 2 and Fig. 3.

The same delta function solutions for $[Y^{0}\Omega^{0}]^{0}$ states can also
be found using the momentum basis in Case~1B by solving the equations in
(\ref{B.1b}). In this case one uses normalized spherical Bessel functions
$N\rho j_{0}(k_{m}\rho)=\sqrt{2/\rho_{0}}\sin(k_{m}\rho)$ where $m=1,2,...,$
$M$ $\ $with $M=40$ and $\rho_{0}=10^{-4}$ $Bohr$ to find the solutions in
Region 1 corresponding to Fig. 3. The calculated anomalous state wave
functions in this basis are shown in Fig. 4. These solutions must be linear
combinations of the free particle states $\Psi_{S}^{0}$ in (\ref{B.1c}) which
have $E=0.$ The radial solutions for these $[Y^{0}\Omega^{0}]^{0}$ anomalous
bound-states are also approximate delta functions. We can write the overall
wave function as%
\begin{equation}
|\Psi_{S}^{0},i\rangle\simeq C\delta(\rho-\rho_{i})Y^{0}\Omega^{0}%
(\boldsymbol{e}_{11}-\boldsymbol{e}_{22})/\sqrt{2}, \label{2.9a}%
\end{equation}
with energy $E_{i}\simeq-e^{2}/\rho_{i}$ where $C$ is the normalization and%

\begin{equation}
\rho_{i}=i\Delta\rho\text{ for }i=1,2,...,M-1, \label{2.10}%
\end{equation}
as in the case of the coordinate representation (\ref{2.8}), (\ref{2.9}) where
the spacing is now
\begin{equation}
\Delta\rho=\rho_{0}/M. \label{2.10a}%
\end{equation}
Again, we note that in Fig. 4 the component $(y_{12}^{\alpha}+y_{21}^{\alpha
})$ is too small to be seen.\ 

Instead of solving the set of equations (\ref{B.1b}), these same anomalous
bound-state solutions can be obtained more directly by solving the equation in
the same momentum basis (denoting $k_{m}$ by $m)$, equivalent to (\ref{x.1}):%

\begin{equation}
-\sum_{m^{\prime}}\left\langle e^{2}/\rho\right\rangle _{mm^{\prime}}%
(c_{11}^{0}-c_{22}^{0})_{m^{\prime}}=E(c_{11}^{0}-c_{22}^{0})_{m} \label{x.2}%
\end{equation}
where $(c_{11}^{0}+c_{22}^{0})_{m}=0$ and $\left\langle -e^{2}/\rho
\right\rangle _{mm^{\prime}}=V_{mm^{\prime}}^{0}$ (see Appendix B)$.$ Again,
the anomalous energies and wave functions are found by simply diagonalizing
the potential matrix $V_{mm^{\prime}}^{0}$. The resulting radial solutions of
(\ref{x.2}) are indistinguishable from the solutions shown in Fig. 4.

In general, for a finite basis set, in either the coordinate or momentum
representation, the solutions to the equation%
\begin{equation}
V(\rho)\psi(\rho)=E\psi(\rho). \label{2.1}%
\end{equation}
are found to be the approximate delta functions such that\textit{ }%
\begin{gather}
\psi^{\prime}(\rho)\simeq C\delta(\rho-\rho^{\prime}),\label{2.2}\\
E^{\prime}\simeq V(\rho^{\prime}),\nonumber
\end{gather}
with $C$ and the discrete grid $\rho^{\prime}$ to be determined. Such
solutions are called the \textit{discrete variable representation (DVR)} and
are found by simply diagonalizing the potential $V(\rho)$ in the chosen
basis\textit{. }They numerically converge to exact delta functions only for an
infinite basis for which they form highly localized bound-states. This assumes
that the basis set is complete and obeys the completeness relation, which for
the spherical Bessel functions is%
\begin{equation}
\frac{2}{\rho_{0}}\sum_{m=1}^{\infty}\rho^{2}j_{L}(k_{m}\rho)j_{L}(k_{m}%
\rho^{\prime})=\delta(\rho-\rho^{\prime})\ \ for\ L=0,1,.... \label{2.3}%
\end{equation}
The DVR wave functions are useful in quantum chemistry because any potential
$V\left(  \rho\right)  $ is diagonal in this basis with energy $V(\rho
^{\prime})$ at the points $\rho^{\prime}$. For any finite basis set, the
discrete variable representation forms a discrete grid, similar to a finite
element grid, at positions $\rho_{i}$ corresponding to the zeros of the basis
functions. The grid then depends on the basis set chosen to form the DVR
representation. For a review of the literature see the work of Light and
Carrington \cite{Light2000} and the references therein.

For a finite basis set, one can find the normalization $C$ of the approximate
delta functions $C\delta(\rho-\rho_{i})$ using the property%

\begin{equation}
\int\delta(\rho-\rho_{i})~d\rho=1. \label{2.4}%
\end{equation}
For a delta function $\delta(\rho-\rho_{i})$ that has a height $\delta(0)$ and
a full-width-at-half-maximum (FWHM) $\Delta\rho,$ the integral becomes%

\[
\int\delta(\rho-\rho_{i})~d\rho=\delta(0)\Delta\rho=1,
\]
or%
\begin{equation}
\delta(0)=1/\Delta\rho. \label{2.5}%
\end{equation}
The normalization condition on the delta function $C\delta(\rho-\rho_{i})$
then becomes%

\[
C^{2}\int\delta(\rho-\rho_{i})^{2}d\rho=C^{2}\delta(0)^{2}\Delta\rho=1,
\]
so that%

\begin{equation}
C=\sqrt{\Delta\rho}. \label{2.6}%
\end{equation}

In the momentum basis, the approximated delta functions for the $i$th
eigenfunction $(y_{11}^{0}-y_{22}^{0})_{i}/\sqrt{2}$ are given simply by the
completeness condition on the radial basis $N\rho j_{0}(k_{m}\rho)$ in
(\ref{2.3}) for a finite basis set at the discrete $\rho_{i}$,%

\begin{align}
(y_{11}^{0}-y_{22}^{0})_{i}/\sqrt{2}  &  =\sqrt{\frac{\rho_{0}}{M}}\frac
{2}{\rho_{0}}\sum_{m=1}^{M-1}\sin(k_{m}\rho)\sin(k_{m}\rho_{i}),\label{3.14}\\
&  \simeq\sqrt{\Delta\rho}\delta(\rho-\rho_{i}),\nonumber
\end{align}
for $i=1,2,...,M-1$. These analytic wave functions for $M=40$ are compared to
those wave functions calculated numerically in Fig. 4. Thus, the completeness
relation allows us to readily derive these analytic DVR solutions for the
momentum basis. One can also show that (\ref{3.14}) forms an orthonormal basis
set such that%
\begin{align}
I_{ij}  &  =%
\frac12
\int_{0}^{\rho_{0}}d\rho(y_{11}^{0}-y_{22}^{0})_{i}(y_{11}^{0}-y_{22}^{0}%
)_{j},\label{3.16}\\
&  =\Delta\rho(\frac{2}{\rho_{0}})^{2}\int_{0}^{\rho_{0}}d\rho\sum_{m=1}%
^{M-1}\sum_{m^{\prime}=1}^{M-1}\sin(k_{m}\rho)\sin(k_{m}\rho_{i}%
)\sin(k_{m^{\prime}}\rho)\sin(k_{m^{\prime}}\rho_{j}),\nonumber\\
&  =\frac{2}{\sqrt{M\rho_{0}}}\sum_{m=1}^{M-1}\sin(k_{m}\rho_{i})\sin
(k_{m}\rho_{j}),\nonumber\\
&  =\delta_{ij}.\nonumber
\end{align}

So far, only the anomalous states $|\Psi_{S}^{0},i\rangle$ in (\ref{2.9a}) for
$[Y^{0}\Omega^{0}]^{0}$ have been considered. However, one obtains the exact
same set of equations for case 3B states with $[Y^{0}\Omega^{0}]^{0}.$ In
particular, for the $\Psi_{A}^{0}$ states in (\ref{B.3c}) one has the equation
in (\ref{B.3b}) after replacing $E$ with $E+e^{2}/\rho,$
\[
-\sum_{m^{\prime}}\left\langle e^{2}/\rho\right\rangle _{mm^{\prime}}%
(c_{12}^{0}-c_{21}^{0})_{m^{\prime}}=E(c_{12}^{0}-c_{21}^{0})_{m}%
\]
where $(c_{12}^{0}+c_{21}^{0})_{m}=0$ and $\left\langle -e^{2}/\rho
\right\rangle _{mm^{\prime}}=V_{mm^{\prime}}^{0}.$ The solutions for these
case~3 anomalous bound-states are%
\begin{equation}
|\Psi_{A}^{0},i\rangle=\sqrt{\Delta\rho}\delta(\rho-\rho_{i})Y^{0}\Omega
^{0}(\boldsymbol{e}_{12}-\boldsymbol{e}_{21})/\sqrt{2}, \label{3.17}%
\end{equation}
with energy $E_{i}=-e^{2}/\rho_{i}.$ Thus, there are two degenerate solutions
$|\Psi_{S}^{0},i\rangle$ and $|\Psi_{A}^{0},i\rangle$ for the $Y^{0}\Omega
^{0}$ anomalous states which have radial wave functions $\sqrt{\Delta\rho
}\delta(\rho-\rho_{i})$ and energies $-e^{2}/\rho_{i}.$

There are also two anomalous bound-state $J=0$ solutions with $[Y^{L}%
\Omega^{S}]^{J}=[Y^{1}\Omega^{1}]^{0}$ arising from $|\Psi_{A}^{\alpha
},i\rangle$ and $|\Psi_{S}^{\alpha},i\rangle$ for case~1B Eq. (\ref{B.1c}) and
case~3B Eq. (\ref{B.3c}), respectively. Summarizing these results, one has the
four $J=0$ anomalous bound-states which encompass the four different possible
$C$ and $P$ parities as denoted in Appendix B:
\begin{equation}%
\begin{tabular}
[c]{cccccc}%
$State$ & $\psi(\rho)[Y^{L}\Omega^{S}]^{J}$ & $Dirac\ Vector$ & $Case$ & $C$ &
$P$\\\cline{2-6}%
$|\Psi_{S}^{0},i\rangle$ & \multicolumn{1}{|c}{$\sqrt{\Delta\rho}\delta
(\rho-\rho_{i})[Y^{0}\Omega^{0}]^{0}$} & \multicolumn{1}{|c}{$(\boldsymbol{e}%
_{11}-\boldsymbol{e}_{22})/\sqrt{2}$} & \multicolumn{1}{|c}{$1$} &
\multicolumn{1}{|c}{$1$} & \multicolumn{1}{|c|}{$-1$}\\
$|\Psi_{A}^{0},i\rangle$ & \multicolumn{1}{|c}{$\sqrt{\Delta\rho}\delta
(\rho-\rho_{i})[Y^{0}\Omega^{0}]^{0}$} & \multicolumn{1}{|c}{$(\boldsymbol{e}%
_{12}-\boldsymbol{e}_{21})/\sqrt{2}$} & \multicolumn{1}{|c}{$3$} &
\multicolumn{1}{|c}{$-1$} & \multicolumn{1}{|c|}{$1$}\\
$|\Psi_{A}^{\alpha},i\rangle$ & \multicolumn{1}{|c}{$\sqrt{\Delta\rho}%
\delta(\rho-\rho_{i})[Y^{1}\Omega^{1}]^{0}$} &
\multicolumn{1}{|c}{$(\boldsymbol{e}_{12}-\boldsymbol{e}_{21})/\sqrt{2}$} &
\multicolumn{1}{|c}{$1$} & \multicolumn{1}{|c}{$-1$} &
\multicolumn{1}{|c|}{$-1$}\\
$|\Psi_{S}^{\alpha},i\rangle$ & \multicolumn{1}{|c}{$\sqrt{\Delta\rho}%
\delta(\rho-\rho_{i})[Y^{1}\Omega^{1}]^{0}$} &
\multicolumn{1}{|c}{$(\boldsymbol{e}_{11}-\boldsymbol{e}_{22})/\sqrt{2}$} &
\multicolumn{1}{|c}{$3$} & \multicolumn{1}{|c}{$1$} & \multicolumn{1}{|c|}{$1$%
}\\\cline{2-6}%
\end{tabular}
\ \ \ \ \ \ \ \ \ \ \ \ \ \ \ \label{3.20}%
\end{equation}
The $|\Psi_{S}^{0},i\rangle$ are the only $J=0$ anomalous states which have
the same $C$ and $P$ parities as the atomic ground-state, namely $C=1$ and
$P=-1$. As a result, they are the only anomalous states which can interact
with the atomic ground-state.

\subsubsection{Coupling of the Atomic Ground State with Anomalous States}

It can now be shown that the anomalous wave functions $|\Psi_{S}^{0},i\rangle$
above for $[Y^{0}\Omega^{0}]^{0}$ are coupled by the Coulomb potential $V_{C}$
with the zero-order Dirac ground-state atomic wave function $|\Psi_{D}%
\rangle^{(0)}$ for $[Y^{0}\Omega^{0}]^{0}$. This coupling explains the unusual
behavior of the Dirac wave functions in Region 1 in Fig. 1. This coupling can
be determined analytically and compared to the Dirac wave functions calculated
in Ref. \cite{Scott1992} which are duplicated in Fig. 1 on a different scale.

The radial Schroedinger ground-state wave function (in atomic units) for
positronium is
\[
y_{S}(\rho)=\frac{\rho e^{-\rho/2}}{\sqrt{2}}.
\]
The Dirac ground-state vector to zero order, is then%
\begin{equation}
|\Psi_{D}\rangle^{(0)}\simeq y_{S}Y^{0}\Omega^{0}\boldsymbol{e}_{11}.
\label{3.16a}%
\end{equation}
The Coulomb coupling of $|\Psi_{D}\rangle^{(0)}$ with anomalous state
$|\Psi_{S}^{0},i\rangle$ at $\rho_{i}$ in (\ref{3.20}) is given simply by the
integral approximation over $\Delta\rho$
\begin{align*}
\langle\Psi_{S}^{0},i|\frac{-e^{2}}{\rho}|\Psi_{D}\rangle^{(0)}  &
=\langle\sqrt{\Delta\rho/2}\delta(\rho-\rho_{i})(\boldsymbol{e}_{11}%
-\boldsymbol{e}_{22})|\frac{-e^{2}}{\rho}|y_{S}(\rho)\boldsymbol{e}%
_{11}\rangle\simeq-\sqrt{\Delta\rho/2}\frac{e^{2}}{\rho_{i}}y_{S}(\rho
_{i})\delta(0)\Delta\rho,\\
&  =-\sqrt{\Delta\rho/2}\frac{e^{2}}{\rho_{i}}y_{S}(\rho_{i}),
\end{align*}
using $\delta(0)\Delta\rho=1$ (\ref{2.5})$.\ $From first order perturbation
theory, the first order correction of the Dirac ground-state $|\Psi_{D}%
\rangle$ of energy $E_{D}\simeq2mc^{2}$ due to the Coulomb coupling with the
anomalous state $|\Psi_{S}^{0},i\rangle$ of energy $E_{i}=-e^{2}/\rho_{i}$ is%
\begin{align*}
|\Psi_{D}(\rho_{i})\rangle^{(1)}  &  =\frac{\langle\Psi_{S}^{0},i|\frac
{-e^{2}}{\rho}|\Psi_{D}\rangle^{(0)}}{E_{D}-E_{i}}|\Psi_{S}^{0},i\rangle,\\
&  =-y_{S}(\rho_{i})\left\{  \frac{e^{2}}{2\rho_{i}}\frac{1}{(2mc^{2}%
+e^{2}/\rho_{i})}\right\}  Y^{0}\Omega^{0}(\boldsymbol{e}_{11}-\boldsymbol{e}%
_{22}).
\end{align*}
where we evaluate $|\Psi_{S}^{0},i\rangle$ at $\rho=\rho_{i}$. For an infinite
basis set, the DVR grid at $\rho_{i}$ becomes continuous. Letting%
\begin{equation}
g=\frac{e^{2}}{2\rho_{i}}\frac{1}{(2mc^{2}+e^{2}/\rho_{i})}, \label{3.18}%
\end{equation}
the positronium ground-state wave function at $\rho_{i}$ becomes, to first
order,
\[
|\Psi_{D}(\rho_{i})\rangle\simeq|\Psi_{D}(\rho_{i})\rangle^{(0)}+|\Psi
_{D}(\rho_{i})\rangle^{(1)}=y_{S}(\rho_{i})(1-g)Y^{0}\Omega^{0}\boldsymbol{e}%
_{11}+y(\rho_{i})gY^{0}\Omega^{0}\boldsymbol{e}_{22}.
\]
The analytic values of the Dirac radial components are then%
\begin{equation}
y_{11}=y_{S}(1-g),\ \ \ y_{22}=y_{S}g. \label{3.19}%
\end{equation}

The square of these Dirac components, $y_{22}^{2}$ and $y_{11}^{2},$ are
plotted in Fig. 5. This very simple result (\ref{3.19}) explains quite well
the unusual behavior of the ground-state Dirac wave function of positronium in
Region 1. Note that the analytic values of $y_{22}^{2}$ are not correct in
Region 3 where the above approximations fail. Assuming an infinite basis set,
the $\rho_{i}$ become continuous $\rho_{i}\rightarrow\rho^{\prime}$ and the
Dirac delta functions become exact $\delta(\rho-\rho^{\prime})$ (\ref{2.3}).
In the limit $\rho_{i}\rightarrow0,$ where $e^{2}/\rho_{i}\gg2mc^{2}$, it can
be seen from (\ref{3.18}) that $g\rightarrow%
\frac12
$ and these Dirac components (\ref{3.19}) become equal to one-half the
Schroedinger wave function, $y_{11}=y_{22}=%
\frac12
y_{S}$. The fact that the two Dirac components $y_{11}^{2}$ and $y_{22}^{2}$
converge for small $\rho$ can be clearly seen in Region~1 of Fig.~5. In
general, for any $(nLSJ)=(n000)$ Schroedinger radial wave function $y_{n}%
(\rho),$ one can obtain the Dirac radial components $y_{11}$ and $y_{22}$ by
replacing $y_{S}$ in (\ref{3.19}) with $y_{n}$ where $y_{1}=y_{S}$.

\subsection{Magnetic Potential}

The atomic bound-states and anomalous bound-states behave quite differently in
the presence of a magnetic potential $\boldsymbol{V}_{M}$. The two-body Dirac
equation is now solved including the magnetic potential so that the new
Hamiltonian is
\begin{equation}
\boldsymbol{H=H}_{0}\boldsymbol{+V}_{C}+\boldsymbol{V}_{M}.
\end{equation}
The form of $\boldsymbol{V}_{M}$ is different for atomic and anomalous states.

\subsubsection{Atomic States}

To determine the fine structure of atomic positronium to order $mc^{2}%
\alpha^{4}$ one must include the magnetic potential. For atomic bound-states
the appropriate potential is in the Coulomb gauge. This is because, for atomic
states, where $\hbar k\ll mc,$ the Dirac operators $\boldsymbol{\alpha}_{e}$
and $\boldsymbol{\alpha}_{p}$ have expectation values close to the fine
structure constant $\alpha=e^{2}/\hslash c\sim v/c$. In the Coulomb gauge one
obtains the Breit potential \cite{Alstine1997},%

\begin{equation}
\boldsymbol{V}_{M}=\boldsymbol{V}_{B}(\rho)=\frac{e^{2}}{2\rho}%
[\boldsymbol{\alpha}_{e}\cdot\boldsymbol{\alpha}_{p}+(\boldsymbol{\alpha}%
_{e}\cdot\boldsymbol{\hat{r}}_{e})(\boldsymbol{\alpha}_{p}\cdot
\boldsymbol{\hat{r}}_{p})],
\end{equation}
as result of second order perturbation theory corresponding to the exchange of
a transverse photon.

For atomic bound-states of positronium, the Breit potential gives fine
structure corrections of order $mc^{2}\alpha^{4}$ which can be determined, as
in the case of the Coulomb potential above, using expectation values
$\left\langle \boldsymbol{V}_{M}\right\rangle $ in the Pauli approximation.
One finds \cite{Bethe1957}, letting $H_{5}^{B}=H_{5}^{B}(1)+H_{5}^{B}(2),$%

\begin{align}
H_{2}^{B}  &  =\frac{mc^{2}\alpha^{4}}{8n^{4}}-\frac{3mc^{2}\alpha^{4}}%
{8n^{3}(2L+1)}+\frac{mc^{2}\alpha^{4}}{8n^{3}}\delta_{L0},\\
H_{3}^{B}  &  =\frac{2mc^{2}\alpha^{4}}{8n^{3}L(L+1)(2L+1)}\left\{
\begin{tabular}
[c]{c}%
$L$\\
$-1$\\
$-(L+1)$%
\end{tabular}%
\begin{tabular}
[c]{c}%
$\text{ }J=L+1$\\
$\text{ }J=L$\\
$\text{ }J=L-1$%
\end{tabular}
\ \ \right\}  \text{ }(1-\delta_{L0})\delta_{S1},\nonumber\\
H_{5}^{B}(1)  &  =-\frac{mc^{2}\alpha^{4}}{4n^{3}}\delta_{S0}\delta_{L0}%
+\frac{mc^{2}\alpha^{4}}{12n^{3}}\delta_{S1}\delta_{L0},\nonumber\\
H_{5}^{B}(2)  &  =-\frac{mc^{2}\alpha^{4}}{8n^{3}L(L+1)(2L+1)}\left\{
\begin{tabular}
[c]{c}%
$L/(2L+3)$\\
$-1$\\
$(L+1)/(2L-1)$%
\end{tabular}%
\begin{tabular}
[c]{c}%
$\text{ }J=L+1$\\
$\text{ }J=L$\\
$\text{ }J=L-1$%
\end{tabular}
\ \ \right\}  \text{ }(1-\delta_{L0})\delta_{S1}.\nonumber
\end{align}
To this order, there is an additional term due to the energy change resulting
from positronium annihilation for $S=1$ states given by%

\begin{equation}
H_{an}=\frac{mc^{2}\alpha^{4}}{4n^{3}}\delta_{S1}\delta_{L0}.
\end{equation}
The Pauli energies $E_{p}^{\prime}$ of $\boldsymbol{H}$ including $H_{an}$ is then,%

\begin{equation}
E_{P}^{\prime}=2mc^{2}+H_{0}+H_{1}+H_{3}^{C}+H_{4}^{C}+H_{2}^{B}+H_{3}%
^{B}+H_{5}^{B}+H_{an}. \label{3.21}%
\end{equation}
Combining terms, one finds to fourth order \cite{Ferrell1951}%

\begin{align}
E_{P}^{\prime}  &  =2mc^{2}-\frac{mc^{2}\alpha^{2}}{4n^{2}}+\frac{mc^{2}%
\alpha^{4}}{n^{3}}\{\frac{11}{64n}-\frac{1}{2(2L+1)}+\xi\delta_{S1}\},\\
\text{where }\xi &  =\frac{7\delta_{L0}}{12}+\frac{1-\delta_{L0}}%
{4(2L+1)}\left\{
\begin{tabular}
[c]{c}%
$\frac{3L+4}{(L+1)(2L+3)}$\\
$-\frac{1}{L(L+1)}$\\
$-\frac{3L-1}{L(2L-1)}$%
\end{tabular}%
\begin{tabular}
[c]{c}%
$\text{ }J=L+1$\\
$\text{ }J=L$\\
$\text{ }J=L-1$%
\end{tabular}
\right\}  \text{ .}\nonumber
\end{align}
Note that $H_{3}=H_{3}^{C}+H_{3}^{B}$ has both a Coulomb term and a Breit term .

In Table 2 the Coulomb energy corrections, $E_{C}=H_{3}^{C}+H_{4}^{C},$ are
compared with the Breit energy corrections, $E_{B}=H_{2}^{B}+H_{3}^{B}%
+H_{5}^{B}.$ Also shown in Table 2 are the total Pauli energies $E_{P}%
^{\prime}-2mc^{2}$ (\ref{3.21}) in $Hartree$ which can be compared to
$E_{P}-2mc^{2}$ (\ref{1.26}) in Table 1. This comparison allows one to
determine the electric and magnetic contributions to the fine structure. It is
important to realize that the Breit energy correction is found from the
expectation value of the Breit potential for the various states and not by a
diagonalization as was the case for the Coulomb potential. This means that if
one adds the energy corrections for the Breit terms and $H_{an}$ to both
$E_{D}$ and $E_{P}$ in Table 1, the differences $E_{D}-E_{P}$ remain the same
to order $mc^{2}\alpha^{6}$ . Table 1 and Table 2 allow one to understand the
two-body Dirac equation in the Coulomb gauge in the context of positronium spectroscopy.%

\begin{table}[tbp] \centering
\begin{tabular}
[c]{|c|c|cc}\hline
$n$\textbf{\ }$L$\textbf{\ }$S$\textbf{\ }$J$ &
\multicolumn{1}{|c|}{$\boldsymbol{E}_{C}$\textbf{\ }$(10^{-9})$} &
\multicolumn{1}{|c|}{$\boldsymbol{E}_{B}$\textbf{\ }$(10^{-9})$} &
\multicolumn{1}{c|}{$\boldsymbol{E}_{P}^{\prime}\boldsymbol{-2mc}^{2}$%
}\\\hline
1 0 0 0 & \multicolumn{1}{|r|}{2 497. 469 23} & \multicolumn{1}{|r}{-19 979.
753 85} & \multicolumn{1}{|c|}{-0.250 017 482 284 62}\\
1 0 1 1 & \multicolumn{1}{|r|}{2 497. 469 23} & \multicolumn{1}{|r}{\ -2 219.
972 65} & \multicolumn{1}{|c|}{-0.249 986 402 667 52}\\
2 0 0 0 & \multicolumn{1}{|r|}{\ \ 156. 091 83} & \multicolumn{1}{|r}{\ -2
913. 714 10} & \multicolumn{1}{|c|}{-0.062 502 757 622 28}\\
2 0 1 1 & \multicolumn{1}{|r|}{\ 156. 091 83} &
\multicolumn{1}{|r}{\ \ \ \ -693. 741 45} & \multicolumn{1}{|c|}{-0.062 498
872 670 14}\\
2 1 0 1 & \multicolumn{1}{|r|}{-121. 75 404} & \multicolumn{1}{|r}{-416. 244
87} & \multicolumn{1}{|c|}{-0.062 500 537 649 62}\\
2 1 1 0 & \multicolumn{1}{|r|}{-398. 901 34} & \multicolumn{1}{|r}{-1248. 734
62} & \multicolumn{1}{|c|}{-0.062 501 647 635 95}\\
2 1 1 1 & \multicolumn{1}{|r|}{-260. 153 45} & \multicolumn{1}{|r}{-554. 993
16} & \multicolumn{1}{|c|}{-0.062 500 815 146 21}\\
2 1 1 2 & \multicolumn{1}{|r|}{\ \ \ 17. 343 54} & \multicolumn{1}{|r}{-166.
497 95} & \multicolumn{1}{|c|}{-0.062 500 149 154 41}\\
3 0 0 0 & \multicolumn{1}{|r|}{\ \ \ 30. 832 95} & \multicolumn{1}{|r}{-904.
433 30} & \multicolumn{1}{|c|}{-0.027 778 651 378 13}\\
3 0 1 1 & \multicolumn{1}{|r|}{\ \ 30. 832 95} & \multicolumn{1}{|r}{-246. 663
63} & \multicolumn{1}{|c|}{-0.027 777 500 281 20}\\
3 1 0 1 & \multicolumn{1}{|r|}{\ -51. 388 26} & \multicolumn{1}{|r}{-164. 442
42} & \multicolumn{1}{|c|}{-0.027 777 993 608 45}\\
3 1 1 0 & \multicolumn{1}{|r|}{-133. 609 47} & \multicolumn{1}{|r}{-411. 106
05} & \multicolumn{1}{|c|}{-0.027 778 322 493 29}\\
3 1 1 1 & \multicolumn{1}{|r|}{\ -92. 498 86} & \multicolumn{1}{|r}{-205. 553
02} & \multicolumn{1}{|c}{-0.027 778 075 829 66}\\
3 1 1 2 & \multicolumn{1}{|r|}{\ -10. 277 65} & \multicolumn{1}{|r}{-90. 443
33} & \multicolumn{1}{|c|}{-0.027 777 878 498 76}\\
3 2 0 2 & \multicolumn{1}{|r|}{\ -18. 499 77} & \multicolumn{1}{|r}{-65. 776
97} & \multicolumn{1}{|c|}{-0.027 777 862 054 52}\\
3 2 1 1 & \multicolumn{1}{|r|}{\ -43. 166 13} & \multicolumn{1}{|r}{-123. 331
81} & \multicolumn{1}{|c|}{-0.027 777 944 275 73}\\
3 2 1 2 & \multicolumn{1}{|r|}{-26. 721 89} & \multicolumn{1}{|r}{-73. 999 09}
& \multicolumn{1}{|c|}{-0.027 777 878 498 76}\\
3 2 1 3 & \multicolumn{1}{|r|}{\ \ -2. 055 53} & \multicolumn{1}{|r}{-35. 237
66} & \multicolumn{1}{|c|}{-0.027 777 815 070 97}\\\hline
\end{tabular}
\caption{Coulomb, Breit, and Total Pauli Energies in Hartree}\label{Table 2}%
\end{table}%

\subsubsection{Anomalous States}

The energies of the anomalous bound-states are considerably shifted when the
magnetic potential is included in the Hamiltonian. The magnetic part of the
potential can be evaluated much more easily for the anomalous states than for
the atomic states. However, the Breit potential cannot be used for the
anomalous bound-states because the magnetic potential is as strong as the
Coulomb potential. In fact, for the anomalous state delta functions
$\delta(\rho-\rho^{\prime})$, which have very high momentum $\hbar k\gg mc$,
the Dirac operators $\boldsymbol{\alpha}_{e}$ and $\boldsymbol{\alpha}_{p}$
have expectation values of $v/c=1$ and not $v/c\sim\alpha$ as in the case for
atomic states. For such high momentum states, one must now use the covariant
Feynman gauge instead of the Coulomb gauge. For this gauge, one uses the Gaunt
potential $\boldsymbol{V}_{G}$ \cite{Alstine1997} instead of the Breit
potential $V_{B}$ where%

\begin{equation}
\boldsymbol{V}_{M}(\rho)=\boldsymbol{V}_{G}(\rho)=\frac{e^{2}}{\rho
}\boldsymbol{\alpha}_{e}\cdot\boldsymbol{\alpha}_{p},
\end{equation}
or, equivalently,%

\begin{equation}
\boldsymbol{V}_{G}(\rho)=\frac{e^{2}}{\rho}\left(
\begin{tabular}
[c]{cccc}%
$0$ & $0$ & $0$ & $\boldsymbol{\sigma}_{e}\boldsymbol{\cdot\sigma}_{p}\text{
}$\\
$0$ & $0$ & $\boldsymbol{\sigma}_{e}\boldsymbol{\cdot\sigma}_{p}\text{ }$ &
$0$\\
$0$ & $\boldsymbol{\sigma}_{e}\boldsymbol{\cdot\sigma}_{p}\text{ }$ & $0$ &
$0$\\
$\boldsymbol{\sigma}_{e}\boldsymbol{\cdot\sigma}_{p}\text{ }$ & $0$ & $0$ &
$0$%
\end{tabular}
\ \ \ \ \ \ \ \ \ \ \ \right)  .
\end{equation}
For the singlet and triplet spin functions $\Omega_{0}^{0}$ and $\Omega
_{\Sigma}^{1}$, respectively, one finds%
\begin{align}
(\boldsymbol{\sigma}_{e}\boldsymbol{\cdot\sigma}_{p})\Omega_{0}^{0}  &
=-3\Omega_{0}^{0},\label{4.18}\\
(\boldsymbol{\sigma}_{e}\boldsymbol{\cdot\sigma}_{p})\Omega_{\Sigma}^{1}  &
=\Omega_{\Sigma}^{1}.\nonumber
\end{align}
such that%
\begin{align}
(\boldsymbol{\alpha}_{e}\cdot\boldsymbol{\alpha}_{p})\Omega_{0}^{0}%
(\boldsymbol{e}_{11}-\boldsymbol{e}_{22})  &  =3\Omega_{0}^{0}(\boldsymbol{e}%
_{11}-\boldsymbol{e}_{22}),\label{4.19}\\
(\boldsymbol{\alpha}_{e}\cdot\boldsymbol{\alpha}_{p})\Omega_{0}^{0}%
(\boldsymbol{e}_{12}-\boldsymbol{e}_{21})  &  =3\Omega_{0}^{0}(\boldsymbol{e}%
_{12}-\boldsymbol{e}_{21}),\nonumber\\
(\boldsymbol{\alpha}_{e}\cdot\boldsymbol{\alpha}_{p})\Omega_{\Sigma}%
^{1}(\boldsymbol{e}_{11}-\boldsymbol{e}_{22})  &  =-\Omega_{\Sigma}%
^{1}(\boldsymbol{e}_{11}-\boldsymbol{e}_{22}),\nonumber\\
(\boldsymbol{\alpha}_{e}\cdot\boldsymbol{\alpha}_{p})\Omega_{\Sigma}%
^{1}(\boldsymbol{e}_{12}-\boldsymbol{e}_{21})  &  =-\Omega_{\Sigma}%
^{1}(\boldsymbol{e}_{12}-\boldsymbol{e}_{21}).\nonumber
\end{align}
Importantly, the anomalous bound-states are eigenfunctions of the total
potentials $V_{C},~V_{G},~$and $E=V_{C}+V_{G}$. From (\ref{4.19}) and
(\ref{3.20}), the eigenvalues of the potentials $V_{C},~V_{G},~$and
$E=V_{C}+V_{G}$ are \ given in (\ref{4.20}).%

\begin{equation}%
\begin{tabular}
[c]{cccccc}%
$State$ & $\psi(\rho)[Y^{L}\Omega^{S}]^{J}$ & $Dirac\ Vector$ & $V_{C}$ &
$V_{G}$ & $E=V_{C}+V_{G}$\\\cline{2-6}%
$|\Psi_{S}^{0},i\rangle$ & \multicolumn{1}{|c}{$\sqrt{\Delta\rho}\delta
(\rho-\rho_{i})[Y^{0}\Omega^{0}]^{0}$} & \multicolumn{1}{|c}{$(\boldsymbol{e}%
_{11}-\boldsymbol{e}_{22})/\sqrt{2}$} & \multicolumn{1}{|c}{$-\frac{e^{2}%
}{\rho_{i}}$} & \multicolumn{1}{|c}{$\frac{3e^{2}}{\rho_{i}}$} &
\multicolumn{1}{|c|}{$\frac{2e^{2}}{\rho_{i}}$}\\
$|\Psi_{A}^{0},i\rangle$ & \multicolumn{1}{|c}{$\sqrt{\Delta\rho}\delta
(\rho-\rho_{i})[Y^{0}\Omega^{0}]^{0}$} & \multicolumn{1}{|c}{$(\boldsymbol{e}%
_{12}-\boldsymbol{e}_{21})/\sqrt{2}$} & \multicolumn{1}{|c}{$-\frac{e^{2}%
}{\rho_{i}}$} & \multicolumn{1}{|c}{$\frac{3e^{2}}{\rho_{i}}$} &
\multicolumn{1}{|c|}{$\frac{2e^{2}}{\rho_{i}}$}\\
$|\Psi_{A}^{\alpha},i\rangle$ & \multicolumn{1}{|c}{$\sqrt{\Delta\rho}%
\delta(\rho-\rho_{i})[Y^{1}\Omega^{1}]^{0}$} &
\multicolumn{1}{|c}{$(\boldsymbol{e}_{12}-\boldsymbol{e}_{21})/\sqrt{2}$} &
\multicolumn{1}{|c}{$-\frac{e^{2}}{\rho_{i}}$} & \multicolumn{1}{|c}{$-\frac
{e^{2}}{\rho_{i}}$} & \multicolumn{1}{|c|}{$-\frac{2e^{2}}{\rho_{i}}$}\\
$|\Psi_{S}^{\alpha},i\rangle$ & \multicolumn{1}{|c}{$\sqrt{\Delta\rho}%
\delta(\rho-\rho_{i})[Y^{1}\Omega^{1}]^{0}$} &
\multicolumn{1}{|c}{$(\boldsymbol{e}_{11}-\boldsymbol{e}_{22})/\sqrt{2}$} &
\multicolumn{1}{|c}{$-\frac{e^{2}}{\rho_{i}}$} & \multicolumn{1}{|c}{$-\frac
{e^{2}}{\rho_{i}}$} & \multicolumn{1}{|c|}{$-\frac{2e^{2}}{\rho_{i}}$%
}\\\cline{2-6}%
\end{tabular}
\ \ \ \ \ \ . \label{4.20}%
\end{equation}

It is interesting that there is a doublet for the anomalous states instead of
a singlet or a triplet as in the case of atomic states. In the case of
anomalous bound-states, the effective mass $mc^{2}=\pm2e^{2}/\rho_{i}$
\ depends on the spin components and the degeneracy depends on the Dirac
components. This is the reverse of the atomic case. One can also view the
doublets $|\Psi_{A}^{\alpha},i\rangle$ and $|\Psi_{S}^{\alpha},i\rangle$ to be
the antiparticles of the doublets $|\Psi_{S}^{0},i\rangle$ and $|\Psi_{A}%
^{0},i\rangle$ because they have the opposite mass. Furthermore, the delta
functions $\delta(\rho-\rho_{i})$ for the anomalous bound-states
$|\Psi,i\rangle$ show that the electron and positron cannot overlap and
annihilate. The lack of overlap between $\delta(\rho-\rho_{i})$ and
$\delta(\rho-\rho_{j})$ also means that there can be no radiative transitions
between anomalous states $|\Psi,i\rangle$ and $|\Psi,j\rangle$. That is, for
$\rho_{i}\neq\rho_{j}$, there are no multipole moments between anomalous
bound-states $|\Psi,i\rangle$ and $|\Psi,j\rangle$ where
\[
\left\langle \delta(\rho-\rho_{i})\left\vert \rho^{n}\right\vert \delta
(\rho-\rho_{j})\right\rangle =0.
\]

\section{Bethe-Salpeter Equations for Positronium: Separability of Atomic and
Anomalous States}

It is now shown that the Bethe-Salpeter equation insures the complete
separability between the atomic and anomalous states. That is, in the presence
of potentials $V_{C}$ and $V_{M}$ above, the atomic and anomalous states
cannot interact. Although we treat only $V_{C}$ here, the same result applies
to $V_{M}.$ This means that the Dirac wave functions in Fig. 1 and Fig. 5 are
incorrect because the Pauli and anomalous wave functions are erroneously
coupled by the Coulomb potential $V_{C}$ in the two-body Dirac equation. Only
the Pauli wave functions in Fig. 1 are correct near the origin for the order calculated.

The two-body Dirac equation for positronium is only an approximation to the
Bethe-Salpeter equation \cite{Salpeter1951} which is relativistically
invariant and can include all the necessary QED (quantum electrodynamic)
corrections. Ultimately, any justification for using the two-body Dirac
equation for positronium comes from the Bethe-Salpeter equation. Furthermore,
the Bethe-Salpeter equation itself has been related to S-matrix field theory
by Gell-Mann and Low \cite{Gell-Mann1951} and Sucher \cite{Sucher1957} to
justify its application to bound-states. Thus, the Bethe-Salpeter equation may
be thought of as equivalent to bound-state QED, whereas the original S-matrix
field theory was applied to scattering QED. More recent treatments of
bound-state QED include works by Sapirstein and Yennie \cite{Sapirstein1990},
Ito \cite{Ito97}, and Grant \cite{Grant2007}. However, one must keep in mind
that the anomalous bound-states can only occur for equal mass atoms such as
positronium and not for the general hydrogenic atom originally considered by
Salpeter and others. In this section, natural units are used where $c=$
$\hslash=1.$

It is useful to define the times $\tau$ and $T$ as the fourth components of
the relative coordinate four vectors $\rho\equiv(\boldsymbol{\rho},i\tau)$ and
$R\equiv(\boldsymbol{R},iT)$, respectively in (\ref{1.4}). One can also define
the energies $\varepsilon$ and $E$ as the fourth components of the conjugate
momentum four vectors $\pi\equiv(\boldsymbol{\pi},i\varepsilon)$ and
$P\equiv(\boldsymbol{P},iE)$, respectively in (\ref{1.5}). Note that the
notation is changed in this section so that $\rho,$ for example, is now a
four-vector and is not equal to $\left\vert \boldsymbol{\rho}\right\vert $
which will now be written out explicitly. From (\ref{1.4}) one finds the times%

\begin{equation}
\tau=t_{e}-t_{p},\ \ \ T=%
\frac12
(t_{e}+t_{p}), \label{4.1a}%
\end{equation}
and from (\ref{1.5}) the conjugate energies
\begin{equation}
\varepsilon=%
\frac12
(e_{e}-e_{p}),\ \ \ E=e_{e}+e_{p}, \label{4.1b}%
\end{equation}
in terms of the one-body times and energies. The problem with the two-body
Dirac equation is that its Hamiltonian formalism does not properly treat the
relative time $\tau$ in the advanced or retarded potential nor the relative
energy $\varepsilon$. Indeed, the two-body Dirac equation is only a function
of the time $T$ and energy $E.$ Thus, it is implicit that $\tau=0$ so that
$t_{e}=t_{p}=T$ and $\varepsilon=0$ so that $e_{e}=e_{p}=E/2.$ It has already
been shown that the anomalous states of the two-body Dirac equation are
coupled to the atomic states by the Coulomb potential. This coupling is
responsible for the fact that $y_{11}^{2}=y_{22}^{2}$ near the origin in Fig.
5. The Bethe-Salpeter equation is now used for a Coulomb potential, treating
the relative time and energy $\tau$ and $\varepsilon$ explicitly, in order to
show that these anomalous solutions of positronium are actually uncoupled from
the atomic solutions.

The Feynman derivation of QED for positronium \cite{Feynman1949} is based on
the one-body Green's function for the one-body Dirac equation. Similarly the
Bethe-Salpeter equation is based on the two-body Green's function for the
two-body Dirac equation. The two-body Dirac equation for positronium with a
Coulomb potential is only accurate to order $m\alpha^{4}$ \cite{Salpeter1952}.
Indeed Karplus and Klein \cite{Karplus1952} and Fulton and Martin
\cite{Fulton1954} have given the corrections to order $m\alpha^{5}$ for
positronium using the Bethe-Salpeter equation \cite{Salpeter1951}%
,\cite{Salpeter1952}. This equation is now used to derive the atomic and
anomalous solutions to positronium in a Coulomb potential, which differ
significantly from those of the two-body Dirac equation.

The two-body Green's function for the electron and positron is simply the
product of the one-body Green's functions $K(r_{e},r_{e}^{\prime}%
)K(r_{p},r_{p}^{\prime})\equiv K^{2}$ and is a solution of the two-body equation,%

\begin{equation}
\lbrack i(\gamma^{e}\cdot p_{e})+m_{e}][i(\gamma^{p}\cdot p_{p})+m_{p}%
]K^{2}=\delta^{4}(r_{e}-r_{e}^{\prime})\delta^{4}(r_{p}-r_{p}^{\prime}),
\label{4.9}%
\end{equation}
without stipulating the boundary conditions. Transforming to the relative
coordinates in the momentum representation with $\boldsymbol{P}=\boldsymbol{0}%
$ where $\boldsymbol{\pi=p}_{e}\boldsymbol{=-p}_{p}$ and using
$\boldsymbol{\alpha=}i\beta\boldsymbol{\gamma}$, $\beta=\gamma_{4\text{ }}$one
finds%
\begin{equation}
K^{2}=-\frac{\gamma_{4}^{e}}{\left[  \varepsilon+%
\frac12
E-\boldsymbol{h}_{0}^{e}\right]  }\frac{\gamma_{4}^{p}}{\left[  \varepsilon-%
\frac12
E+\boldsymbol{h}_{0}^{p}\right]  }, \label{4.10}%
\end{equation}
where%
\begin{align*}
\boldsymbol{h}_{0}^{e}\psi_{\pm}^{e}(\boldsymbol{\pi})  &  =\pm e_{0}\psi
_{\pm}^{e}(\boldsymbol{\pi}),\\
\boldsymbol{h}_{0}^{p}\psi_{\pm}^{p}(-\boldsymbol{\pi})  &  =\pm e_{0}%
\psi_{\pm}^{p}(-\boldsymbol{\pi})\\
e_{0}  &  =\sqrt{\pi^{2}+m^{2}.}%
\end{align*}
The Coulomb potential $V_{C}(\rho)$ in the momentum representation is%
\[
G_{C}(-\boldsymbol{k)}=-\frac{1}{(2\pi)^{3}}\int d^{3}\rho
\ e^{i\boldsymbol{k\cdot\rho}}\frac{e^{2}}{\rho}=-\frac{e^{2}}{2\pi^{2}}%
\frac{1}{\boldsymbol{k\cdot k}}.
\]
The two-body Bethe-Salpeter equation in the momentum representation for the
Coulomb potential becomes%
\begin{align}
\Psi(\boldsymbol{\pi)}  &  =-\frac{\gamma_{4}^{e}\gamma_{4}^{p}}{2\pi i}\int
d^{3}k\ \int_{-\infty}^{\infty}d\varepsilon\ K^{2}G_{C}(-\boldsymbol{k)}%
\Psi(\boldsymbol{k+\pi}),\label{4.11}\\
&  =\frac{1}{2\pi i}\int d^{3}k\int_{-\infty}^{\infty}d\varepsilon\ \frac
{1}{\left[  \varepsilon+%
\frac12
E-\boldsymbol{h}_{0}^{e}\right]  }\frac{1}{\left[  \varepsilon-%
\frac12
E+\boldsymbol{h}_{0}^{p}\right]  }G_{C}(-\boldsymbol{k})\Psi(\boldsymbol{k+\pi
}),\nonumber
\end{align}
where it is assumed, for simplicity, that $\Psi$ is independent of energy
$\varepsilon$ although an equivalent equation occurs under more general
conditions \cite{Salpeter1952}. Projecting with operators $\boldsymbol{\Lambda
}_{\pm\pm}$\ on (\ref{4.11}), where
\begin{align*}
\boldsymbol{\Lambda}_{\pm\pm}  &  \boldsymbol{=}\boldsymbol{\Lambda}_{\pm}%
^{e}(\boldsymbol{\pi)\Lambda}_{\pm}^{p}(-\boldsymbol{\pi)}=|\psi_{\pm}%
^{e}(\boldsymbol{\pi)}\psi_{\pm}^{p}(-\boldsymbol{\pi)}\rangle\langle\psi
_{\pm}^{e}(\boldsymbol{\pi)}\psi_{\pm}^{p}(-\boldsymbol{\pi)|,}\\
&  =\frac{e_{0}\pm\boldsymbol{h}_{0}^{e}(\boldsymbol{\pi)}}{2e_{0}}\frac
{e_{0}\pm\boldsymbol{h}_{0}^{p}(\boldsymbol{\pi)}}{2e_{0}},
\end{align*}
one finds%
\begin{align}
\Psi_{\pm\pm}(\boldsymbol{\pi)}  &  =\frac{\boldsymbol{1}}{2\pi i}%
\int_{-\infty}^{\infty}d\varepsilon\frac{1}{\left[  \varepsilon-\varepsilon
_{e}\right]  }\frac{1}{\left[  \varepsilon-\varepsilon_{p}\right]  }%
\Gamma_{\pm\pm}(\boldsymbol{\pi)},\label{4.12}\\
\Gamma_{\pm\pm}(\boldsymbol{\pi)}  &  =\boldsymbol{\Lambda}_{\pm\pm}\int
d^{3}kG(-\boldsymbol{k})\Psi(\boldsymbol{k+\pi}),\nonumber
\end{align}
where $K^{2}$ has poles at
\[
\varepsilon_{e}=-%
\frac12
E\pm e_{0},\ \ \ \varepsilon_{p}=%
\frac12
E\mp e_{0},
\]
corresponding to the first and second terms in the denominator of
(\ref{4.10}), respectively, when operating on $\psi_{\pm}^{e}(\boldsymbol{\pi
)}\psi_{\pm}^{p}(-\boldsymbol{\pi)}$. \ The wave functions $\Psi_{\pm\pm
}(\boldsymbol{\pi)}$ are now evaluated below where it is shown that
\textbf{one must use the Feynman propagator }$K=K_{F}$\textbf{ for all atomic
states and the Retarded propagator }$K=K_{R}$\textbf{ for anomalous
bound-states}.

For both $K_{F}$ and $K_{R}$ one chooses $e=e_{0}-i\epsilon$ so that
positive-energy states $\psi_{+}$ propagate forward in time. The question
becomes: how should negative-energy states $\psi_{-}$ propagate in time? For
the Feynman single particle propagator $K_{F}$ one chooses $e=-e_{0}%
+i\epsilon$ so that negative-energy states $\psi_{-}$ propagate backward in
time and for the Retarded single particle propagator $K_{R}$ one chooses
$e=-e_{0}-i\epsilon$ so that negative-energy states $\psi_{-}$ propagate
forward in time. So one has the following results for the time propagation of
positive- and negative-energy states:%
\begin{align}
&  K_{F}\text{\ }\left\{
\begin{tabular}
[c]{c}%
$\psi_{+}$:$\ \ e=+e_{0}-i\epsilon$, \ \ \ forward\ \\
$\psi_{-}$:$\ \ e=-e_{0}+i\epsilon\text{, \ backward\ }$%
\end{tabular}
\ \ \right\}  \text{all atomic states},\label{4.8}\\
&  K_{R}\left\{  \text{\ }%
\begin{tabular}
[c]{c}%
$\psi_{+}$:$\ \ e=+e_{0}-i\epsilon$, \ \ \ forward\ \\
$\psi_{-}$:$\ \ e=-e_{0}-i\epsilon\text{, \ \ forward}$%
\end{tabular}
\ \ \right\}  \text{anomalous bound-states}.\nonumber
\end{align}

The correct temporal boundary condition on the negative-energy states
$\psi_{-}$ depends on physical consistency with the known temporal behavior of
the electron or positron undergoing scattering or annihilation. For free
particles, both negative-energy electrons and positrons must propagate
backward in time for two important reasons \cite{Feynman1949,Sakurai1967}.
First, if a negative-energy particle propagated forward in time, then a
positive-energy particle could scatter into a negative-energy particle and be
lost at a later time. This is not possible because of particle conservation.
Second, a virtual negative-energy electron must propagate backward in time so
that it can annihilate with a virtual positive-energy electron which is moving
forward in time. That is, a virtual negative-energy electron moving backward
in time is equivalent to a positron moving forward in time. Such time behavior
accounts for virtual electron-positron pairs. This boundary condition
$e=-e_{0}+i\epsilon$ for negative-energy states $\psi_{-}$ of free particles
is determined from physical reasoning although both boundary conditions are
mathematically allowed. Thus, for $K$ to correspond to propagation of free
electrons or positrons, one must choose $e=\pm e_{0}\mp i\epsilon$ for the
proper boundary conditions corresponding to $K_{F}$. We show below that this
choice of $K_{F}$ also must also apply to atomic bound-states. However, it
will also be shown below that $K_{F}$ cannot be used for electrons or
positrons in anomalous bound-states where the particles are not free.

\subsubsection{Atomic State Propagator $K_{F}^{2}$}

One can now use the temporal boundary conditions for $K=K_{F}$ in (\ref{4.8}).
Consider first the poles of $\Psi_{++}(\boldsymbol{\pi)}$ in (\ref{4.12}) at
\[
\varepsilon_{e}=-%
\frac12
E+e_{0}-i\epsilon,\ \ \ \varepsilon_{p}=%
\frac12
E-e_{0}+i\epsilon.
\]
One can complete the line integral in either the upper or lower half complex
plane where the integrand is convergent with the same results%
\begin{align*}
\Psi_{++}(\boldsymbol{\pi)}  &  =\frac{1}{2\pi i}\int_{-\infty}^{\infty
}d\varepsilon\text{ }\frac{1}{[\varepsilon-\varepsilon_{e}][\varepsilon
-\varepsilon_{p}]}\Gamma_{++}(\boldsymbol{\pi)},\\
&  =\frac{1}{E-2e_{0}}\Gamma_{++}(\boldsymbol{\pi)}.
\end{align*}
Similarly, for the poles of $\Psi_{--}(\boldsymbol{\pi)}$ in (\ref{4.12}) at%
\[
\varepsilon_{e}=-%
\frac12
E-e_{0}+i\epsilon,\ \ \ \varepsilon_{p}=%
\frac12
E+e_{0}-i\epsilon,
\]
the integral becomes%
\begin{align*}
\Psi_{--}(\boldsymbol{\pi)}  &  =\frac{1}{2\pi i}\int_{-\infty}^{\infty
}d\varepsilon\text{ }\frac{1}{[\varepsilon-\varepsilon_{e}][\varepsilon
-\varepsilon_{p}]}\Gamma_{--}(\boldsymbol{\pi)},\\
&  =-\frac{1}{E+2e_{0}}\Gamma_{--}(\boldsymbol{\pi)}.
\end{align*}
Finally, for the $K_{F}$ propagator, one finds for the corresponding integrals
in (\ref{4.12}) that $\Psi_{+-}(\boldsymbol{\pi)}=\Psi_{-+}(\boldsymbol{\pi
)}=0$. Combining these results and letting
\[
\boldsymbol{\Lambda}=\boldsymbol{\Lambda}_{++}(\boldsymbol{\pi}%
)-\boldsymbol{\Lambda}_{--}(\boldsymbol{\pi}),
\]
one has%
\[
(E-2e_{0})\Psi_{_{++}}(\boldsymbol{\pi})+(E+2e_{0})\Psi_{--}(\boldsymbol{\pi
})=\boldsymbol{\Lambda}\Gamma(\boldsymbol{\pi}),
\]
or%
\begin{align}
\lbrack\boldsymbol{h}_{0}^{e}(\boldsymbol{\pi})+\boldsymbol{h}_{0}%
^{p}(\boldsymbol{\pi})][\Psi_{_{++}}(\boldsymbol{\pi})+\Psi_{--}%
(\boldsymbol{\pi})]-\frac{e^{2}}{2\pi^{2}}\boldsymbol{\Lambda}\int d^{3}%
k\frac{1}{\boldsymbol{k\cdot k}}\Psi(\boldsymbol{k+\pi})  &  =E[\Psi_{_{++}%
}(\boldsymbol{\pi})+\Psi_{--}(\boldsymbol{\pi})],\label{4.13}\\
\Psi_{+-}(\boldsymbol{\pi)}  &  =\Psi_{-+}(\boldsymbol{\pi)}=0,\nonumber
\end{align}
which is the Bethe-Salpeter equation for the atomic states for a Coulomb
potential in the ladder approximation.

Because of the operator $\boldsymbol{\Lambda,}$ the Coulomb potential is
attractive for positive-energy atomic states but is repulsive for
negative-energy atomic states. Thus the positive and negative wave functions
have bound-states with the opposite energies. Only the free particle states
$\Psi_{_{++}}(\boldsymbol{\pi})$ and $\Psi_{--}(\boldsymbol{\pi})$ with
energies $E=2e_{0}$ and $E=-2e_{0},$ respectively, contribute to the atomic
states for a Coulomb potential to all orders in the ladder approximation.

One can compare the above equation to the two-body Dirac equation in the
momentum representation%

\begin{equation}
\lbrack\boldsymbol{h}_{0}^{e}(\boldsymbol{\pi})+\boldsymbol{h}_{0}%
^{p}(\boldsymbol{\pi})]\Psi(\boldsymbol{\pi})-\frac{e^{2}}{2\pi^{2}}\int
d^{3}k\frac{1}{\boldsymbol{k\cdot k}}\Psi(\boldsymbol{k+\pi})=E\Psi
(\boldsymbol{\pi}), \label{4.14}%
\end{equation}
which erroneously includes the anomalous states $\Psi_{+-}(\boldsymbol{\pi})$
and $\Psi_{-+}(\boldsymbol{\pi}).$ The solutions to (\ref{4.14}) in the
coordinate representation lead to the Dirac solutions shown in Fig. 1 while
the solutions to (\ref{4.13}) lead to the Pauli solutions shown in Fig. 1 (to
the calculated order). The Dirac solutions in Fig. 1 and Fig. 5 are incorrect
because they include Coulomb coupling between the atomic and anomalous
bound-states which cannot occur.

\subsubsection{Anomalous State Propagator $K_{R}^{2}$}

The anomalous bound-states which include the $\Psi_{+-}(\boldsymbol{\pi})$ and
$\Psi_{-+}(\boldsymbol{\pi})$ wave functions are indeed solutions of the
Bethe-Salpeter equation if one changes the boundary conditions. The
appropriate boundary conditions for anomalous states correspond to both
positive- and negative-energy states propagating forward in time $\tau$
corresponding to the two-body Retarded propagator $K_{R}^{2}$. That is, one
can now use the temporal boundary conditions for $K=K_{R}$ in (\ref{4.8}).
This means that the negative-energy states which comprise the anomalous
bound-states cannot exist as separate free particles: negative-energy states
of free particles must propagate backward in time because they correspond to
the antiparticle propagating forward in time.

But this is not a problem for anomalous bound-states. Probability is still
conserved for bound-states despite the fact that negative- and positive-energy
states can scatter into each other. Further, there can be no electron-positron
annihilation for such states as shown in Sec. III. This means that the
discrete variable representation (DVR) is necessary to produce the correct
boundary conditions for the proper $K_{R}^{2}$ propagation. It appears that
anomalous bound-states are both mathematically and physically allowed.

With this proper time behavior, the poles of $\Psi_{+-}(\boldsymbol{\pi})$ in
(\ref{4.12}) are
\[
\varepsilon_{e}=-%
\frac12
E+e_{0}-i\epsilon,\ \ \ \varepsilon_{p}=%
\frac12
E+e_{0}+i\epsilon.
\]
The integral for the Bethe-Salpeter equation becomes%
\[
\Psi_{+-}(\boldsymbol{\pi})=\frac{1}{E}\Gamma_{+-}(\boldsymbol{\pi}),
\]
where one may close the line integral in either the upper- or lower-half
complex plane. Similarly, the poles of $\Psi_{-+}(\boldsymbol{\pi})$ in
(\ref{4.12}) are%
\[
\varepsilon_{e}=-%
\frac12
E-e_{0}-i\epsilon,\ \ \ \varepsilon_{p}=%
\frac12
E-e_{0}+i\epsilon,
\]
and the integral becomes%
\[
\Psi_{-+}(\boldsymbol{\pi})=\frac{1}{E}\Gamma_{-+}(\boldsymbol{\pi}).
\]
Finally, for the $K_{R}$ propagator, one finds for the corresponding integrals
in (\ref{4.12}) that $\Psi_{++}(\boldsymbol{\pi})=\Psi_{--}(\boldsymbol{\pi
})=0$. Combining these results, one has%
\begin{align}
-\frac{e^{2}}{2\pi^{2}}[\boldsymbol{\Lambda}_{+-}(\boldsymbol{\pi
})+\boldsymbol{\Lambda}_{-+}(\boldsymbol{\pi})]\int d^{3}k\frac{1}%
{\boldsymbol{k\cdot k}}\Psi(\boldsymbol{k+\pi})  &  =E[\Psi_{_{+-}%
}(\boldsymbol{\pi})+\Psi_{-+}(\boldsymbol{\pi})],\label{4.15}\\
\Psi_{_{++}}(\boldsymbol{\pi})  &  =\Psi_{--}(\boldsymbol{\pi})=0,\nonumber
\end{align}
which is the Bethe-Salpeter equation for the anomalous bound-states for a
Coulomb potential in the ladder approximation. The solutions to this equation
lead to the anomalous bound-state energies shown in Fig. 2 and wave functions
shown in Figs. 3 and 4. As a result, the anomalous bound-states formed from
$\Psi_{+-}(\boldsymbol{\pi})$ and $\Psi_{-+}(\boldsymbol{\pi})$ arise from the
ladder terms of QED and never couple with atomic states. Finally, note that,
for atomic and anomalous bound-states, the Bethe-Salpeter equation leaves no
choice in the temporal boundary conditions but rather they are determined
automatically by this equations.

\section{Conclusions}

It has been shown that there are two types of bound-state solutions to the
two-body Dirac equation and Bethe-Salpeter equation for positronium: there are
the normal atomic solutions and the anomalous solutions. The energies and wave
functions for these two solution have been derived by solving both the
two-body Dirac equation and the Bethe-Salpeter equation with an
electromagnetic potential. The anomalous bound-states wave functions are Dirac
delta functions in the radial coordinate corresponding to the discrete
variable representation (DVR). For these highly localized wave functions, the
electron and positron can neither radiate nor annihilate.

It has also been shown that the numerical accuracy of the atomic Dirac
energies are of order $mc^{2}\alpha^{6}$ or less when compared to the analytic
Pauli energies. The Dirac bound-state wave function radial components,
however, differs significantly near the origin from their Pauli approximations
and are incorrect. This difference is because the Dirac equation erroneously
couples the atomic ground-state wave function with the anomalous bound-state
wave functions near the origin due to the electromagnetic potential. No such
coupling can occurs for the Bethe-Salpeter equation because of the different
time behavior for the atomic and anomalous negative-energy states.

Finally, it has been shown that one must use the Feynman two-body propagator
$K_{F}^{2}$ for the atomic bound-states. On the other hand, one must use the
Retarded two-body propagator $K_{R}^{2}$ for the anomalous bound-states.
Unlike atomic states, anomalous states can never be free. For the atomic
bound-states, the free particle states are useful where the momentum is
quantized. On the other hand, for the anomalous bound-states, the discrete
variable representation (DVR) must be used where the position is quantized. 

\appendix

\section{Coordinate Representation for Two-Body Dirac Equations}

These equations are given for a free particle in spherical coordinates. For a
Coulomb potential replace the energy $E$ by $E-V_{C}$ below where
$V_{C}=-e^{2}/\rho$. We define the recoupling coefficients $a$ and $b$ as in
(\ref{1.12}). The three cases 1A, 2A, and 3A of equations below are in
agreement with those of Malenfant (see Refs. \cite{Malenfant1988} and
\cite{Scott1992}) for sets 1, 3, and 2, respectively using the explicitly
symmetrized basis for the wave function coefficients. Note that the radial
functions for all cases, $y_{ij}(\rho),$ include the radial scale factor
$\rho$ for which one has the boundary condition $y_{ij}(0)=0.$The three
different cases below are labeled by their dominant $\Psi_{11}$ component for
the atomic states.

\subsection{Case 1A: $S=0,$ $L=J$.}

For the basis,%
\begin{equation}
\Psi=\frac{1}{\rho}\left(
\begin{array}
[c]{c}%
y_{11}^{0}(\rho)[Y^{J}\Omega^{0}]_{N}^{J}\\
i\{y_{12}^{+}(\rho)[Y^{J+1}\Omega^{1}]_{N}^{J}+y_{12}^{-}(\rho)[Y^{J-1}%
\Omega^{1}]_{N}^{J}\}\\
i\{y_{21}^{+}(\rho)[Y^{J+1}\Omega^{1}]_{N}^{J}+y_{21}^{-}(\rho)[Y^{J-1}%
\Omega^{1}]_{N}^{J}\}\\
y_{22}^{0}(\rho)[Y^{J}\Omega^{0}]_{N}^{J}%
\end{array}
\right)  , \label{A.1a}%
\end{equation}
one has, using (\ref{1.11})-(\ref{1.14}),%
\begin{align}
2mc^{2}(y_{11}^{0}-y_{22}^{0})-2a\hbar c[\frac{d}{d\rho}+\frac{J+1}{\rho
}](y_{12}^{+}+y_{21}^{+})+2b\hbar c[\frac{d}{d\rho}-\frac{J}{\rho}](y_{12}%
^{-}+y_{21}^{-})  &  =E(y_{11}^{0}+y_{22}^{0}),\label{A.1b}\\
2mc^{2}(y_{11}^{0}+y_{22}^{0})  &  =E(y_{11}^{0}-y_{22}^{0}),\nonumber\\
2a\hbar c[\frac{d}{d\rho}-\frac{J+1}{\rho}](y_{11}^{0}+y_{22}^{0})  &
=E(y_{12}^{+}+y_{21}^{+}),\nonumber\\
-2b\hbar c[\frac{d}{d\rho}+\frac{J}{\rho}](y_{11}^{0}+y_{22}^{0})  &
=E(y_{12}^{-}+y_{21}^{-}),\nonumber\\
0  &  =E(y_{12}^{+}-y_{21}^{+}),\nonumber\\
0  &  =E(y_{12}^{-}-y_{21}^{-}).\nonumber
\end{align}
The last two equations above are mathematically allowed but have the opposite
charge-conjugation parity from the first four coupled equations. As a result,
these last two equations are uncoupled from the first four equations.

\subsection{Case 2A: $S=1,$ $L=J$.}

For the basis%
\begin{equation}
\Psi=\frac{1}{\rho}\left(
\begin{array}
[c]{c}%
y_{11}^{1}(\rho)[Y^{J}\Omega^{1}]_{N}^{J}\\
i\{y_{12}^{+}(\rho)[Y^{J+1}\Omega^{1}]_{N}^{J}+y_{12}^{-}(\rho)[Y^{J-1}%
\Omega^{1}]_{N}^{J}\}\\
i\{y_{21}^{+}(\rho)[Y^{J+1}\Omega^{1}]_{N}^{J}+y_{21}^{-}(\rho)[Y^{J-1}%
\Omega^{1}]_{N}^{J}\}\\
y_{22}^{1}(\rho)[Y^{J}\Omega^{1}]_{N}^{J}%
\end{array}
\right)  , \label{A.2a}%
\end{equation}

one has the equations%
\begin{align}
2mc^{2}(y_{11}^{1}+y_{22}^{1})+2b\hbar c[\frac{d}{d\rho}+\frac{J+1}{\rho
}](y_{12}^{+}-y_{21}^{+})+2a\hbar c[\frac{d}{d\rho}-\frac{J}{\rho}](y_{12}%
^{-}-y_{21}^{-})  &  =E(y_{11}^{1}-y_{22}^{1}),\label{A.2b}\\
2mc^{2}(y_{11}^{1}-y_{22}^{1})  &  =E(y_{11}^{1}+y_{22}^{1}),\nonumber\\
-2b\hbar c[\frac{d}{d\rho}-\frac{J+1}{\rho}](y_{11}^{1}+y_{22}^{1})  &
=E(y_{12}^{+}-y_{21}^{+}),\nonumber\\
-2a\hbar c[\frac{d}{d\rho}+\frac{J}{\rho}](y_{11}^{1}+y_{22}^{1})  &
=E(y_{12}^{-}-y_{21}^{-}),\nonumber\\
0  &  =E(y_{12}^{+}+y_{21}^{+}),\nonumber\\
0  &  =E(y_{12}^{-}+y_{21}^{-}).\nonumber
\end{align}

\subsection{Case 3A: $S=1,$ $L\neq J$.}

\qquad Case 3A can be obtained from Case 1A and Case 2A by a simple
transformation by using the wave function%

\begin{equation}
\Psi=\frac{1}{\rho}\left(
\begin{array}
[c]{c}%
i\{y_{11}^{+}(\rho)[Y^{J+1}\Omega^{1}]_{N}^{J}+y_{11}^{-}(\rho)[Y^{J-1}%
\Omega^{1}]_{N}^{J}\}\\
y_{12}^{0}(\rho)[Y^{J}\Omega^{0}]_{N}^{J}+y_{12}^{1}(\rho)[Y^{J}\Omega
^{1}]_{N}^{J}\\
y_{21}^{0}(\rho)[Y^{J}\Omega^{0}]_{N}^{J}+y_{21}^{1}(\rho)[Y^{J}\Omega
^{1}]_{N}^{J}\\
i\{y_{22}^{+}(\rho)[Y^{J+1}\Omega^{1}]_{N}^{J}+y_{22}^{-}(\rho)[Y^{J-1}%
\Omega^{1}]_{N}^{J}\}
\end{array}
\right)  . \label{A.3a}%
\end{equation}
Note that the large-large component $\Psi_{11}$ can now correspond to the
atomic state with either $L=J+1$ or $L=J-1.$ The new equations can be found
from Case 1A and Case 2A by the exchange $m_{e}\leftrightarrow-m_{e},$
$y_{11}\leftrightarrow y_{21,}$ $y_{22}\leftrightarrow y_{12},$ or,
equivalently, the exchange $m_{p}\leftrightarrow-m_{p},$ $y_{11}%
\leftrightarrow y_{12,}$ $y_{22}\leftrightarrow y_{21}.$ The two-body Dirac
equation in this basis gives the eight equations for the radial functions (of
which six are coupled),%
\begin{align}
-2a\hbar c[\frac{d}{d\rho}+\frac{J+1}{\rho}](y_{11}^{+}+y_{22}^{+})+2b\hbar
c[\frac{d}{d\rho}-\frac{J}{\rho}](y_{11}^{-}+y_{22}^{-})  &  =E(y_{12}%
^{0}+y_{21}^{0}),\label{A.3b}\\
2mc^{2}(y_{11}^{+}-y_{22}^{+})+2a\hbar c[\frac{d}{d\rho}-\frac{J+1}{\rho
}](y_{12}^{0}+y_{21}^{0})  &  =E(y_{11}^{+}+y_{22}^{+}),\nonumber\\
2mc^{2}(y_{11}^{-}-y_{22}^{-})-2b\hbar c[\frac{d}{d\rho}+\frac{J}{\rho
}](y_{12}^{0}+y_{21}^{0})  &  =E(y_{11}^{-}+y_{11}^{-}),\nonumber\\
0  &  =E(y_{12}^{0}-y_{21}^{0}),\nonumber\\
2b\hbar c[\frac{d}{d\rho}+\frac{J+1}{\rho}](y_{11}^{+}-y_{22}^{+})+2a\hbar
c[\frac{d}{d\rho}-\frac{J}{\rho}](y_{11}^{-}-y_{22}^{-})  &  =E(y_{12}%
^{1}-y_{21}^{1}),\nonumber\\
2mc^{2}(y_{11}^{+}+y_{22}^{+})-2b\hbar c[\frac{d}{d\rho}-\frac{J+1}{\rho
}](y_{12}^{1}-y_{21}^{1})  &  =E(y_{11}^{+}-y_{22}^{+}),\nonumber\\
2mc^{2}(y_{11}^{-}+y_{22}^{-})-2a\hbar c[\frac{d}{d\rho}+\frac{J}{\rho
}](y_{12}^{1}-y_{21}^{1})  &  =E(y_{11}^{-}-y_{11}^{-}),\nonumber\\
0  &  =E(y_{12}^{1}+y_{21}^{1}).\nonumber
\end{align}
With this transformation, the two separate sets of three coupled equations in
the Case 1A and Case 2A basis now become six coupled equations in Case 3. The
two uncoupled equations are allowed but have opposite charge-conjugation
parity from the other equations. The charge-conjugation \ and inversion parity
is shown clearly in Appendix B.

\section{Momentum Representation for Two-Body Dirac Equations}

Refer to (\ref{1.18}), (\ref{1.19}) for the definitions of the states
$|L,S,k\rangle$ given below. As in the case of the coordinate representation,
there are three Cases 1B, 2B, and 3B of equations in the momentum
representation. This representation is given for the free particle basis in
spherical coordinates. Below, the energies $E_{\pm\pm}$ for the wave functions
$\Psi_{\pm\pm}$ correspond to%
\begin{align*}
E_{++}  &  =+2e,\ E_{--}=-2e,\\
E_{+-}  &  =E_{-+}=0,
\end{align*}
where%
\[
e=\sqrt{(\hbar ck)^{2}+(mc^{2})^{2}}%
\]
as in (\ref{1.7a}) and (\ref{1.7b}). The anomalous wave functions $\Psi_{+-}%
$and $\Psi_{-+}$ for energies $E_{+-}$ and $E_{-+}$ are symmetrized such that%
\begin{align*}
\Psi_{S}  &  =(\Psi_{+-}+\Psi_{-+})/\sqrt{2},\\
\Psi_{A}  &  =(\Psi_{+-}-\Psi_{-+})/\sqrt{2}.
\end{align*}

The equations for the free particles are diagonal in $k.$ For a Coulomb
potential one must include the relevant potential matrices of the spherical
Bessel functions for a given $J$ :%
\begin{align*}
V_{kk^{\prime}}^{0}  &  =\left\langle J,0,k\left\vert V_{C}\right\vert
J,0,k^{\prime}\right\rangle =-N_{Jk}N_{Jk^{\prime}}\int_{0}^{\rho_{0}}%
d\rho\ \rho j_{J}(k\rho)j_{J}(k^{\prime}\rho),\\
V_{kk^{\prime}}^{1}  &  =\left\langle J,1,k\left\vert V_{C}\right\vert
J,1,k^{\prime}\right\rangle =V_{kk^{\prime}}^{0},\\
V_{kk^{\prime}}^{+}  &  =\left\langle J+1,1,k\left\vert V_{C}\right\vert
J+1,1,k^{\prime}\right\rangle =-N_{Jk}N_{Jk^{\prime}}\int_{0}^{\rho_{0}}%
d\rho\ \rho j_{J+1}(k\rho)j_{J+1}(k^{\prime}\rho),\\
V_{kk^{\prime}}^{-}  &  =\left\langle J-1,1,k\left\vert V_{C}\right\vert
J-1,1,k^{\prime}\right\rangle =-N_{Jk}N_{Jk^{\prime}}\int_{0}^{\rho_{0}}%
d\rho\ \rho j_{J-1}(k\rho)j_{J-1}(k^{\prime}\rho),\\
V_{kk^{\prime}}^{\alpha}  &  =aV_{kk^{\prime}}^{+}+bV_{kk^{\prime}}^{-},\\
V_{kk^{\prime}}^{\beta}  &  =-bV_{kk^{\prime}}^{+}+aV_{kk^{\prime}}^{-}.
\end{align*}
The energies can then be found by replacing $E\rightarrow E\delta_{kk^{\prime
}}-V_{kk^{\prime}}^{i}$ with the appropriate $i$ and diagonalizing the
resulting matrix for $E$.

The charge-conjugation $C$ parity and inversion $P$ parity of the $\Psi_{++},$
$\Psi_{--},\ \Psi_{S}$, $\Psi_{A}$ states are given below. The $C$ and $P$
parities of the atomic states $\Psi_{++}$, $\Psi_{--}$ are identical to those
given by Malenfant \cite{Malenfant1988} who did not treat the $\Psi_{S}%
,\ \Psi_{A}$ anomalous states. For a given case, only those states with the
same charge-conjugation parity $C$ and $P$ can be coupled by the Coulomb
potential. As in the coordinate representation, the three different cases
below are labeled by their dominant $\Psi_{11}$ component for the atomic states.

\subsection{Case 1B: $S=0,$ $L=J$}

Letting%
\begin{equation}
\Psi=\frac{1}{\rho}\left(
\begin{array}
[c]{c}%
c_{11}^{0}|J,0,k\rangle\\
c_{12}^{\alpha}|J\alpha,1,k\rangle\\
c_{21}^{\alpha}|J\alpha,1,k\rangle\\
c_{22}^{0}|J,0,k\rangle
\end{array}
\right)  , \label{B.1a}%
\end{equation}
the two-body Dirac equation in the momentum basis, for a given $k,J,$ gives
the following three coupled equations for the symmetrized Bessel coefficients,%

\begin{align}
2mc^{2}(c_{11}^{0}-c_{22}^{0})-2\hbar ck(c_{12}^{\alpha}+c_{21}^{\alpha})  &
=E(c_{11}^{0}+c_{22}^{0}),\label{B.1b}\\
2mc^{2}(c_{11}^{0}+c_{22}^{0})  &  =E(c_{11}^{0}-c_{22}^{0}),\nonumber\\
-2\hbar ck(c_{11}^{0}+c_{22}^{0})  &  =E(c_{12}^{\alpha}+c_{21}^{\alpha
}),\nonumber\\
0  &  =E(c_{12}^{\alpha}-c_{21}^{\alpha}).\nonumber
\end{align}
The last equation is uncoupled because it has different charge-conjugation
parity $C$. One obtains the four orthonormal solutions, $\Psi_{i},$ for each
$k,J$\ given in the columns below%
\begin{equation}%
\begin{tabular}
[c]{ccccc}
& $\sqrt{2}\Psi_{++}^{0}$ & $\sqrt{2}\Psi_{--}^{0}$ & $\sqrt{2}\Psi_{S}^{0}$ &
$\sqrt{2}\Psi_{A}^{\alpha}$\\\cline{2-5}%
$(c_{11}^{0}+c_{22}^{0})$ & \multicolumn{1}{|c}{$1$} & $1$ & $.$ &
\multicolumn{1}{c|}{$.$}\\
$(c_{11}^{0}-c_{22}^{0})$ & \multicolumn{1}{|c}{$\frac{mc^{2}}{e}$} &
$-\frac{mc^{2}}{e}$ & $\frac{\sqrt{2}\hbar ck}{e}$ & \multicolumn{1}{c|}{$.$%
}\\
$(c_{12}^{\alpha}+c_{21}^{\alpha})$ & \multicolumn{1}{|c}{$-\frac{\hbar ck}%
{e}$} & $\frac{\hbar ck}{e}$ & $\frac{\sqrt{2}mc^{2}}{e}$ &
\multicolumn{1}{c|}{$.$}\\
$(c_{12}^{\alpha}-c_{21}^{\alpha})$ & \multicolumn{1}{|c}{$.$} & $.$ & $.$ &
\multicolumn{1}{c|}{$\sqrt{2}$}\\\cline{2-5}%
$C$ & $(-1)^{J}$ & $(-1)^{J}$ & $(-1)^{J}$ & $(-1)^{J+1}$\\
$P$ & $(-1)^{J+1}$ & $(-1)^{J+1}$ & $(-1)^{J+1}$ & $(-1)^{J+1}$%
\end{tabular}
\ . \label{B.1c}%
\end{equation}
$\ \ \ \ \ $ The atomic states are labelled $\Psi_{++}^{\lambda}$ and
$\Psi_{--}^{\lambda}$ with superscript $\lambda$ which corresponds to the
dominant component $c_{ij}^{\lambda}\neq0$ \ where $mc^{2}\gg\hbar ck$ and the
anomalous states $\Psi_{S}^{\lambda}$ and $\Psi_{A}^{\lambda}$ are labelled
with superscript $\lambda$ which corresponds to the dominant component
$c_{ij}^{\lambda}\neq0$ where $\hbar ck\gg mc^{2}$. Note that $(c_{11}%
^{0}+c_{22}^{0})=0$ for the anomalous states $\Psi_{S}^{0}$ as in
(\ref{x.2})$.$

\subsection{Case 2B: $S=1,$ $L=J$}

Letting%

\begin{equation}
\Psi=\frac{1}{\rho}\left(
\begin{array}
[c]{c}%
c_{11}^{1}|J,1,k\rangle\\
c_{12}^{\beta}|J\beta,1,k\rangle\\
c_{12}^{\beta}|J\beta,1,k\rangle\\
c_{22}^{1}|J,1,k\rangle
\end{array}
\right)  , \label{B.2a}%
\end{equation}
the two-body Dirac equation in the momentum basis, for a given $k,J,$ gives
the following three coupled equations for the symmetrized Bessel coefficients,%

\begin{align}
2mc^{2}(c_{11}^{1}+c_{22}^{1})-2\hbar ck(c_{12}^{\beta}-c_{21}^{\beta})  &
=E(c_{11}^{1}-c_{22}^{1}),\label{B.2b}\\
2mc^{2}(c_{11}^{1}-c_{22}^{1})  &  =E(c_{11}^{1}+c_{22}^{1}),\nonumber\\
-2\hbar ck(c_{11}^{1}-c_{22}^{1})  &  =E(c_{12}^{\beta}-c_{21}^{\beta
}),\nonumber\\
0  &  =E(c_{12}^{\beta}+c_{21}^{\beta}).\nonumber
\end{align}
The last equation is uncoupled because it has different charge-conjugation
parity $C$. One obtains the four orthonormal solutions $\Psi_{i}$ for each
$k,J$ in the columns below%
\begin{equation}%
\begin{tabular}
[c]{ccccc}
& $\sqrt{2}\Psi_{++}^{1}$ & $\sqrt{2}\Psi_{--}^{1}$ & $\sqrt{2}\Psi_{S}^{1}$ &
$\sqrt{2}\Psi_{A}^{\beta}$\\\cline{2-5}%
$(c_{11}^{1}-c_{22}^{1})$ & \multicolumn{1}{|c}{$1$} & $1$ & $.$ &
\multicolumn{1}{c|}{$.$}\\
$(c_{11}^{1}+c_{22}^{1})$ & \multicolumn{1}{|c}{$\frac{mc^{2}}{e}$} &
$-\frac{mc^{2}}{e}$ & $\frac{\sqrt{2}\hbar ck}{e}$ & \multicolumn{1}{c|}{$.$%
}\\
$(c_{12}^{\beta}-c_{21}^{\beta})$ & \multicolumn{1}{|c}{$-\frac{\hbar ck}{e}$}
& $\frac{\hbar ck}{e}$ & $\frac{\sqrt{2}mc^{2}}{e}$ & \multicolumn{1}{c|}{$.$%
}\\
$(c_{12}^{\beta}+c_{21}^{\beta})$ & \multicolumn{1}{|c}{.} & . & . &
\multicolumn{1}{c|}{$\sqrt{2}$}\\\cline{2-5}%
$C$ & $(-1)^{J+1}$ & $(-1)^{J+1}$ & $(-1)^{J+1}$ & $(-1)^{J}$\\
$P$ & $(-1)^{J+1}$ & $(-1)^{J+1}$ & $(-1)^{J+1}$ & $(-1)^{J+1}$%
\end{tabular}
\ \ \ \ .\ \ \label{B.2c}%
\end{equation}

\subsection{Case 3B: $S=1,$ $L\neq J$}

Letting%
\begin{equation}
\Psi=\frac{1}{\rho}\left(
\begin{array}
[c]{c}%
c_{11}^{\alpha}|J\alpha,1,k\rangle+c_{11}^{\beta}|J\beta,1,k\rangle\\
c_{12}^{0}|J,0,k\rangle+c_{12}^{1}|J,1,k\rangle\\
c_{21}^{0}|J,0,k\rangle+c_{21}^{1}|J,1,k\rangle\\
c_{22}^{\alpha}|J\alpha,1,k\rangle+c_{22}^{\beta}|J\beta,1,k\rangle
\end{array}
\right)  , \label{B.3a}%
\end{equation}
the two-body Dirac equation in the momentum basis, for a given $k,J,$ gives
the following coupled equations for the symmetrized Bessel coefficients,%

\begin{align}
2mc^{2}(c_{11}^{\alpha}-c_{22}^{\alpha})-2\hbar ck(c_{12}^{0}+c_{21}^{0})  &
=E(c_{11}^{\alpha}+c_{22}^{\alpha}),\label{B.3b}\\
2mc^{2}(c_{11}^{\alpha}+c_{22}^{\alpha})  &  =E(c_{11}^{\alpha}-c_{22}%
^{\alpha}),\nonumber\\
-2\hbar ck(c_{11}^{\alpha}+c_{22}^{\alpha})  &  =E(c_{12}^{0}+c_{21}%
^{0}),\nonumber\\
2mc^{2}(c_{11}^{\beta}+c_{22}^{\beta})-2\hbar ck(c_{12}^{1}-c_{21}^{1})  &
=E(c_{11}^{\beta}-c_{22}^{\beta}),\nonumber\\
2mc^{2}(c_{11}^{\beta}-c_{22}^{\beta})  &  =E(c_{11}^{\beta}+c_{22}^{\beta
}),\nonumber\\
-2\hbar ck(c_{11}^{\beta}-c_{22}^{\beta})  &  =E(c_{12}^{1}-c_{21}%
^{1}),\nonumber\\
0  &  =E(c_{12}^{0}-c_{21}^{0}),\nonumber\\
0  &  =E(c_{12}^{1}+c_{21}^{1}).\nonumber
\end{align}
The first six equations consist of two sets of three coupled equations. The
last two equations are uncoupled because they have different
charge-conjugation parity $C$. As in the coordinate representation, these
equations can be found from Case 1B and Case 2B by the exchange $m_{e}%
\longleftrightarrow-m_{e},$ $c_{11}\longleftrightarrow c_{21,}$ $c_{22}%
\longleftrightarrow c_{12},$ or, equivalently, the exchange $m_{p}%
\longleftrightarrow-m_{p},$ $c_{11}\longleftrightarrow c_{12,}$ $c_{22}%
\longleftrightarrow c_{21}$. One obtains the eight orthonormal solutions
$\Psi_{i}$ for each $k,J$ in the columns below%
\begin{equation}%
\begin{tabular}
[c]{ccccccccc}
& $\sqrt{2}\Psi_{++}^{\alpha}$ & $\sqrt{2}\Psi_{--}^{\alpha}$ & $\sqrt{2}%
\Psi_{S}^{\alpha}$ & $\sqrt{2}\Psi_{++}^{\beta}$ & $\sqrt{2}\Psi_{--}^{\beta}$
& $\sqrt{2}\Psi_{A}^{\prime\beta}$ & $\sqrt{2}\Psi_{A}^{0}$ & $\sqrt{2}%
\Psi_{S}^{\prime1}$\\\cline{2-9}%
$(c_{11}^{\alpha}+c_{22}^{\alpha})$ & \multicolumn{1}{|c}{$1$} & $1$ & $.$ &
$.$ & $.$ & $.$ & $.$ & \multicolumn{1}{c|}{$.$}\\
$(c_{11}^{\alpha}-c_{22}^{\alpha})$ & \multicolumn{1}{|c}{$\frac{mc^{2}}{e}$}
& $-\frac{mc^{2}}{e}$ & $\frac{\sqrt{2}\hbar ck}{e}$ & $.$ & $.$ & $.$ & $.$ &
\multicolumn{1}{c|}{$.$}\\
$(c_{12}^{0}+c_{21}^{0})$ & \multicolumn{1}{|c}{$-\frac{\hbar ck}{e}$} &
$\frac{\hbar ck}{e}$ & $\frac{\sqrt{2}mc^{2}}{e}$ & $.$ & $.$ & $.$ & $.$ &
\multicolumn{1}{c|}{$.$}\\
$(c_{11}^{\beta}-c_{22}^{\beta})$ & \multicolumn{1}{|c}{$.$} & $.$ & $.$ & $1$
& $1$ & $.$ & $.$ & \multicolumn{1}{c|}{$.$}\\
$(c_{11}^{\beta}+c_{22}^{\beta})$ & \multicolumn{1}{|c}{$.$} & $.$ & $.$ &
$\frac{mc^{2}}{e}$ & $-\frac{mc^{2}}{e}$ & $\frac{\sqrt{2}\hbar ck}{e}$ & $.$
& \multicolumn{1}{c|}{$.$}\\
$(c_{12}^{1}-c_{21}^{1})$ & \multicolumn{1}{|c}{$.$} & $.$ & $.$ &
$-\frac{\hbar ck}{e}$ & $\frac{\hbar ck}{e}$ & $\frac{\sqrt{2}mc^{2}}{e}$ &
$.$ & \multicolumn{1}{c|}{$.$}\\
$(c_{12}^{0}-c_{21}^{0})$ & \multicolumn{1}{|c}{$.$} & $.$ & $.$ & $.$ & $.$ &
$.$ & $\sqrt{2}$ & \multicolumn{1}{c|}{$.$}\\
$(c_{12}^{1}+c_{21}^{1})$ & \multicolumn{1}{|c}{$.$} & $.$ & $.$ & $.$ & $.$ &
$.$ & $.$ & \multicolumn{1}{c|}{$\sqrt{2}$}\\\cline{2-9}%
$C$ & $(-1)^{J}$ & $(-1)^{J}$ & $(-1)^{J}$ & $(-1)^{J}$ & $(-1)^{J}$ &
$(-1)^{J}$ & $(-1)^{J+1}$ & $(-1)^{J+1}$\\
$P$ & $(-1)^{J}$ & $(-1)^{J}$ & $(-1)^{J}$ & $(-1)^{J}$ & $(-1)^{J}$ &
$(-1)^{J}$ & $(-1)^{J}$ & $(-1)^{J}$%
\end{tabular}
\ \ . \label{B.3c}%
\end{equation}
Here the prime superscript is used for the anomalous wave functions $\Psi
_{A}^{\prime\beta}$ and $\Psi_{S}^{\prime1}$ to distinguish them from their
Case 2B counterparts which have the same $C$ but different $P$. As seen in
(\ref{B.3c}), the two sets of three equations are separable for free particles
just like Case 1B and Case 2B, but will be coupled by the Coulomb potential
because they have the same $C$ and $P$.

\subsection{Addition Theorems.}

Four important addition theorems can be derived for the products of single
particle functions of free particles $g_{n_{e}}^{\ell_{e}\text{ }j_{e}}%
(kr_{e},\theta_{e},\varphi_{e})g_{n_{p}}^{\ell_{p}\text{ }j_{p}}(kr_{p}%
,\theta_{p},\varphi_{p})$ with coordinates $\boldsymbol{r}_{e},\boldsymbol{r}%
_{p}$ where
\begin{align}
g_{n}^{\ell\text{ }j}(kr,\theta,\varphi)  &  \equiv j_{\ell}(kr)[Y^{\ell
}(\theta,\varphi)\chi^{%
\frac12
}]_{n}^{j},\label{B.14.a}\\
&  =j_{\ell}(kr)\sum_{m,\sigma}C_{m\text{\ }\sigma\text{\ }n}^{\ell\text{\ }%
\frac12
\text{\ }j}Y_{m}^{\ell}\chi_{\sigma}^{%
\frac12
},\nonumber
\end{align}
and%
\begin{equation}
\lbrack g_{n_{e}}^{j_{e}\pm%
\frac12
\text{ \ }j_{e}}g_{n_{p}}^{j_{p}\pm%
\frac12
\text{ \ }j_{p}}]_{N}^{J}=\sum_{n_{e},n_{p}}C_{n_{e}\text{\ }n_{p}\text{\ }%
N}^{j_{e}~\text{\ }j_{p}\text{\ }J}g_{n_{e}}^{j_{e}\pm%
\frac12
\text{ \ }j_{e}}g_{n_{p}}^{j_{p}\pm%
\frac12
\text{ \ }j_{p}}. \label{B.14b}%
\end{equation}
The four possible states, $j_{L}(k\rho)[Y^{L}(\theta_{\rho},\varphi_{\rho
})\Omega^{S}]_{N}^{J},$ for a given $J$ with relative coordinates
$\boldsymbol{\rho=r}_{e}-\boldsymbol{r}_{p}$ can be expanded in terms of these
products (\ref{B.14b}). These four addition theorems are derived here from the
work of Danos and Maximon \cite{Danos1965}.

Using the recoupling coefficients $a$ and $b$ in (\ref{1.12}) and ignoring
normalizations $N_{Jk}$ in (\ref{1.18}) and (\ref{1.19}), one obtains%

\begin{align}
|J,0,k\rangle/\rho=  &  j_{J}(k\rho)[Y^{J}(\theta_{\rho},\varphi_{\rho}%
)\Omega^{0}]_{N}^{J}\label{B.10a}\\
&  =\text{ \ }\sum_{j_{e},j_{p}\text{ }(j_{e}-j_{p}-J=even)}q_{j_{e},j_{p}%
:J}\{[g_{n_{e}}^{j_{e}-%
\frac12
\text{ \ }j_{e}}g_{n_{p}}^{j_{p}-%
\frac12
\text{ \ }j_{p}}]_{N}^{J}-[g_{n_{e}}^{j_{e}+%
\frac12
\text{ \ }j_{e}}g_{n_{p}}^{j_{p}+%
\frac12
\text{ \ }j_{p}}]_{N}^{J}\}\nonumber\\
&  \text{ \ \ \ \ \ }-i\sum_{j_{e},j_{p}\text{ }(j_{e}-j_{p}-J=odd)}%
q_{j_{e},j_{p}:J}\{[g_{n_{e}}^{j_{e}-%
\frac12
\text{ \ }j_{e}}g_{n_{p}}^{j_{p}+%
\frac12
\text{ \ }j_{p}}]_{N}^{J}+[g_{n_{e}}^{j_{e}+%
\frac12
\text{ \ }j_{e}}g_{n_{p}}^{j_{p}-%
\frac12
\text{ \ }j_{p}}]_{N}^{J}\},\nonumber
\end{align}
and%

\begin{align}
|J\alpha,1,k\rangle/\rho=i  &  aj_{J+1}(k\rho)[Y^{J+1}(\theta_{\rho}%
,\varphi_{\rho})\Omega^{1}]_{N}^{J}+ibj_{J-1}(k\rho)[Y^{J-1}(\theta_{\rho
},\varphi_{\rho})\Omega^{1}]_{N}^{J}\label{B.10b}\\
&  =i\text{ }\sum_{j_{e},j_{p}\text{ }(j_{e}-j_{p}-J=even)}q_{j_{e},j_{p}%
:J}\{[g_{n_{e}}^{j_{e}-%
\frac12
\text{ \ }j_{e}}g_{n_{p}}^{j_{p}+%
\frac12
\text{ \ }j_{p}}]_{N}^{J}+[g_{n_{e}}^{j_{e}+%
\frac12
\text{ \ }j_{e}}g_{n_{p}}^{j_{p}-%
\frac12
\text{ \ }j_{p}}]_{N}^{J}\}\nonumber\\
&  \text{ \ }-\sum_{j_{e},j_{p}\text{ }(j_{e}-j_{p}-J=odd)}q_{j_{e},j_{p}%
:J}\{[g_{n_{e}}^{j_{e}-%
\frac12
\text{ \ }j_{e}}g_{n_{p}}^{j_{p}-%
\frac12
\text{ \ }j_{p}}]_{N}^{J}-[g_{n_{e}}^{j_{e}+%
\frac12
\text{ \ }j_{e}}g_{n_{p}}^{j_{p}+%
\frac12
\text{ \ }j_{p}}]_{N}^{J}\},\nonumber
\end{align}
where%

\begin{equation}
q_{j_{e},j_{p}:J}=i^{(j_{e}-j_{p}-J)}%
\genfrac{(}{)}{}{}{2\pi\lbrack j_{e}][j_{p}]}{[J]}%
^{%
\frac12
}C_{%
\frac12
\text{\ }-%
\frac12
\text{\ }0}^{j_{e}\text{ \ }j_{p}\text{\ \ }J}. \label{B.10c}%
\end{equation}

One also obtains%

\begin{align}
|J,1,k\rangle/\rho=  &  j_{J}(k\rho)[Y^{J}(\theta_{\rho},\varphi_{\rho}%
)\Omega^{1}]_{N}^{J}\label{B.11a}\\
&  =-\sum_{\text{ }j_{e},j_{p}\text{ }(j_{e}-j_{p}-J=even)}p_{j_{e},j_{p}%
:J}\{[g_{n_{e}}^{j_{e}-%
\frac12
\text{ \ }j_{e}}g_{n_{p}}^{j_{p}-%
\frac12
\text{ \ }j_{p}}]_{N}^{J}+[g_{n_{e}}^{j_{e}+%
\frac12
\text{ \ }j_{e}}g_{n_{p}}^{j_{p}+%
\frac12
\text{ \ }j_{p}}]_{N}^{J}\}\nonumber\\
&  \text{ \ \ \ \ \ \ \ }-i\sum_{j_{e},j_{p}\text{ }(j_{e}-j_{p}%
-J=odd)}p_{j_{e},j_{p}:J}\{[g_{n_{e}}^{j_{e}-%
\frac12
\text{ \ }j_{e}}g_{n_{p}}^{j_{p}+%
\frac12
\text{ \ }j_{p}}]_{N}^{J}-[g_{n_{e}}^{j_{e}+%
\frac12
\text{ \ }j_{e}}g_{n_{p}}^{j_{p}-%
\frac12
\text{ \ }j_{p}}]_{N}^{J}\},\nonumber
\end{align}
and%

\begin{align}
&  |J\beta,1,k\rangle/\rho=-ibj_{J+1}(k\rho)[Y^{J+1}(\theta_{\rho}%
,\varphi_{\rho})\Omega^{1}]_{N}^{J}+iaj_{J-1}(k\rho)[Y^{J-1}(\theta_{\rho
},\varphi_{\rho})\Omega^{1}]_{N}^{J}\label{B.11b}\\
&  =-i\sum_{j_{e},j_{p}\text{ }(j_{e}-j_{p}-J=even)}p_{j_{e},j_{p}%
:J}\{[g_{n_{e}}^{j_{e}-%
\frac12
\text{ \ }j_{e}}g_{n_{p}}^{j_{p}+%
\frac12
\text{ \ }j_{p}}]_{N}^{J}-[g_{n_{e}}^{j_{e}+%
\frac12
\text{ \ }j_{e}}g_{n_{p}}^{j_{p}-%
\frac12
\text{ \ }j_{p}}]_{N}^{J}\}\nonumber\\
&  \text{ \ }-\sum_{\text{ }j_{e},j_{p}\text{ }(j_{e}-j_{p}-J=odd)}%
p_{j_{e},j_{p}:J}\{[g_{n_{e}}^{j_{e}-%
\frac12
\text{ \ }j_{e}}g_{n_{p}}^{j_{p}-%
\frac12
\text{ \ }j_{p}}]_{N}^{J}+[g_{n_{e}}^{j_{e}+%
\frac12
\text{ \ }j_{e}}g_{n_{p}}^{j_{p}+%
\frac12
\text{ \ }j_{p}}]_{N}^{J}\},\nonumber
\end{align}
where%

\begin{equation}
p_{j_{e},j_{p}:J}=i^{(j_{e}-j_{p}-J)}%
\genfrac{(}{)}{}{}{2\pi\lbrack j_{e}][j_{p}]}{[J]}%
^{%
\frac12
}C_{%
\frac12
\text{\ }%
\frac12
\text{\ }1}^{j_{e}\text{\ }j_{p}\text{\ }J}. \label{B.11c}%
\end{equation}

One can readily evaluate the expressions (\ref{B.10a}) and (\ref{B.11a}) for
the special case of $J=0$ so that
\[
j_{0}(k\rho)[Y^{0}(\theta_{\rho},\varphi_{\rho})\Omega^{0}]_{0}^{0}=\sum
_{j}\sqrt{(2\pi\lbrack j]}(-1)^{j-%
\frac12
}\{[g_{n_{e}}^{j-%
\frac12
\text{ \ }j}g_{n_{p}}^{j-%
\frac12
\text{ \ }j}]_{0}^{0}-[g_{n_{e}}^{j+%
\frac12
\text{ \ }j}g_{n_{p}}^{j+%
\frac12
\text{ \ }j}]_{0}^{0}\}\text{,}%
\]
and
\[
j_{1}(k\rho)[Y^{1}(\theta_{\rho},\varphi_{\rho})\Omega^{1}]_{0}^{0}=\sum
_{j}\sqrt{(2\pi\lbrack j]}(-1)^{j-%
\frac12
}\{[g_{n_{e}}^{j+%
\frac12
\text{ \ }j}g_{n_{p}}^{j-%
\frac12
\text{ \ }j}]_{0}^{0}+[g_{n_{e}}^{j-%
\frac12
\text{ \ }j}g_{n_{p}}^{j+%
\frac12
\text{ \ }j}]_{0}^{0}\}\text{.}%
\]
These two special cases have been given previously \cite{Abramowitz1972}.

\section*{Acknowledgments}

The author would like to give special thanks to Drs. Tony Scott, Janine
Shertzer, Gordon Drake, and William Harter. The author has benefitted greatly
from the work of Scott, Shertzer, and Moore \cite{Scott1992} and considers
this work, in many respects, to be an extension of theirs. Drs. Scott and
Shertzer were very helpful in explaining their finite element calculations
which were duplicated here in order to show the anomalous wave functions.
Particular credit goes to Dr. Drake for suggesting the separability between
the anomalous and atomic states. Dr. Harter has been encouraging this work for
a number of years. Thanks are also due to Dr. Joel Kress for extending the
hospitality of Theoretical Division, Los Alamos National Laboratory and to
Drs. Arthur Voter, Brian Kendrick, Peter Miloni, and Lee Collins for many
helpful suggestions during my time at Los Alamos. Dr. Kendrick suggested the
pertinence of the discrete variable representation for the anomalous states.
Finally, the author would like to thank Drs. Robert Sang and Max Standage for
their help and support while at Griffith University.\ 

\section*{References}

\bibliographystyle{apsrev}
\bibliography{chrisref}

\bigskip

\end{document}


\preprint{HEP/123-qed}
\title[Short title for running header]{REV\TeX 4-1}
\author{First Author}
\affiliation{My Institution}
\author{Second Author}
\affiliation{My Institution}
\author{Third Author}
\affiliation{Other Institution}
\keywords{one two three}
\pacs{PACS number}

\begin{abstract}

\end{abstract}
\volumeyear{year}
\volumenumber{number}
\issuenumber{number}
\eid{identifier}
\date[Date text]{date}
\received[Received text]{date}

\revised[Revised text]{date}

\accepted[Accepted text]{date}

\published[Published text]{date}

FIGURE CAPTIONS

\bigskip

FIG. 1. \ The Dirac and Pauli radial wavefunction components (squared) for the
ground-state are compared. Large dots correspond to the 399 grid points used
for the finite element calculation. Plots are on a log-log scale.

\bigskip

FIG. 2. \ The energies $E_{i}$ for the anomalous states are found by solving
the same equations as in Fig.\ 1. The finite element grid points $\rho_{i}$
shown here are the same as in Region 1 of Fig.\ 1. Energies are given
analytically by $-1/\rho_{i}$\ $Hartree$ shown by the solid line. Verticle
lines are at the element boundaries.

\bigskip

FIG. 3. \ The anomalous bound state wavefunctions in the coordinate
representation for the energies in Fig.\ 2 at the element boundaries. The
wavefunctions are approximately evenly spaced delta functions. The component
$(y_{12}^{+}+y_{21}^{+})$ is too small to be seen.

\bigskip

FIG. 4. \ Anomalous bound-state wavefunctions as in Fig.\ 3 but calculated in
the momentum representation. The component $(y_{12}^{\alpha}+y_{21}^{\alpha})$
is too small to be seen. Analytic results are from (38).

\bigskip

FIG. 5. \ The Dirac radial wavefunctions in Fig.\ 1 are compared with
analytical results in which the Dirac zero-order ground-state $|\Psi
_{D}\rangle^{(0)}$ is coupled to the anomalous states $|\Psi_{S}^{0},i\rangle$
by the Coulomb potential. Plots are on a log-log scale.